\documentclass[12pt,a4paper]{amsart}

\usepackage{amssymb}
\makeatletter                                            
\renewcommand{\@begintheorem}[2]{                        
\rm \trivlist \item [\hskip \labelsep {\bf #2\ \ #1.}]   
                                }                        
\makeatother                                             

\makeatletter

\DeclareFontFamily{U}{cyr}{}
\DeclareFontShape{U}{cyr}{m}{n}{
  <5> wncyr5 <6> wncyr6 <7> wncyr7 <8> wncyr8 <9> wncyr9 <10->
wncyr10}{}
\DeclareMathAlphabet{\mathcyr}{U}{cyr}{m}{n}

\input cyracc.def
\input epsf

\newcommand{\ZZ}{{\bf Z}}

\newcommand{\RR}{{\bf R}}
\newcommand{\CC}{{\bf C}}

\newcommand{\FF}{{\bf F}}
\newcommand{\HH}{{\bf H}}
\newcommand{\PP}{{\bf P}}

\newcommand{\im}{\mbox{im}}

\DeclareMathOperator{\diag}{diag}
\DeclareMathOperator{\Imm}{Im}

\title{Modular forms and three loop superstring amplitudes}
\author{Sergio L.~Cacciatori}
\address{Dipartimento di Scienze Fisiche e Matematiche, Universit\`a dell'Insubria,
Via Valleggio 11, I-22100 Como, Italia}
\email{sergio.cacciatori@uninsubria.it}
\author{Francesco Dalla~Piazza}
\address{Dipartimento di Scienze Fisiche e Matematiche, Universit\`a dell'Insubria, 
Via Valleggio 11, I-22100 Como, Italia}
\email{f.dallapiazza@uninsubria.it}
\author{Bert van Geemen}
\address{Dipartimento di Matematica, Universit\`a di Milano,
Via Saldini 50, I-20133 Milano, Italia}
\email{geemen@mat.unimi.it}

\begin{document}
\begin{abstract}
We study a proposal of
D'Hoker and Phong for the chiral superstring measure for genus three.
A minor modification of the constraints they impose 
on certain Siegel modular forms leads to a unique solution. 
We reduce the problem of finding these modular forms, which depend on an even spin structure, to finding a modular form of weight 8 on a certain subgroup of the modular group.
An explicit formula for this form, as a polynomial in the even theta constants,
is given. 
We checked that our result is consistent with the vanishing of the cosmological constant. We also verified a conjecture of D'Hoker and Phong on modular forms in genus 3 and 4 using results of Igusa.
\end{abstract}
\maketitle

\section{Introduction}
Formally string theories in the perturbative approach can be formulated using the path integral formalism outlined by Polyakov, and this is the
starting point for the computation of the scattering amplitudes.
The conformal invariance of the string theory forces the amplitude to be invariant under the action of the modular group as was
exploited by Belavin and Knizhnik \cite{belavin_alggeom} who conjectured: ``any multiloop amplitude in any conformal invariant string theory may be
deduced from purely algebraic objects on moduli spaces $M_p$ of Riemann surfaces''. Indeed, for bosonic strings this permitted the computation of
the vacuum to vacuum amplitude up to four loops in terms of modular forms \cite{BKMP}, \cite{M}. For superstrings there are some difficulties:
the presence of fermionic interactions makes the splitting between chiral and antichiral modes hard, 
moreover one needs a covariant way to integrate
out the Grassmannian variables arising from the supersymmetry on the worldsheet.
In a series of articles, D'Hoker and Phong showed that the computation of $g$-loop amplitudes in string perturbation theory is strictly related to the construction of a suitable measure on the super moduli space of genus $g$ super Riemann surfaces.
They also claimed \cite{DP1}, \cite{DP2}, \cite{D'Hoker:1989rk} that the genus $g$ vacuum to vacuum amplitude should take the form
\begin{equation}
\label{ampiezgenh}
\mathcal{A}=\int_{\mathcal{M}_g}
(\det\Imm\tau)^{-5}\sum_{\Delta,\Delta'}c_{\Delta,\Delta'}
{\rm d}\mu[\Delta](\tau)\wedge
\overline{{\rm d}\mu[\Delta'](\tau)},
\end{equation}
where $\Delta$ and $\Delta'$ denote two spin structures (or theta characteristics), $c_{\Delta,\Delta'}$ are
suitable constant phases depending on the details of the model and ${\rm d}\mu[\Delta](\tau)$ is a holomorphic form of maximal rank
 $(3g-3,0)$ on the moduli space ${\mathcal{M}_g}$ 
of genus $g$ Riemann surfaces.
The Riemann surface is represented by its period matrix $\tau$, after a choice of canonical homology basis.
Since the integrand should be independent from the choice of homology basis, it follows that the measure ${\rm d}\mu[\Delta](\tau)$ must transform
covariantly under the modular group $Sp(2g,\ZZ)$.

In \cite{DPg2a} and following papers, D'Hoker and Phong explicitly solved the problems outlined before for the two loop vacuum to vacuum amplitude, giving
an explicit expression for the two loop  measure in terms of theta constants. Next, in \cite{DP1}, \cite{DP2}, they tried to extend their results
to three loop amplitudes. Mimicking the structure of the two loop chiral measure, they proposed three reasonable constraints (see below) which should
characterize the modular forms composing the measure. 
Then, they tried to find such modular forms, without success.
This negative result, apparently,
can be imputed to their requirement that the modular form, of weight eight, 
should be a product
of the fourth power of a theta constant and modular form of weight six.
This led us to look for a weaker form of the
constraints, in particular relaxing the second one,
by {\em not} requiring such a decomposition and we
do succeed in finding such a form.
Our assumptions are consistent with the expression for the amplitudes at genus one and two, and at genus three they provide
a unique solution.
In this paper we will show the existence, provide an explicit expression for the measure and show that the corresponding cosmological
constant is zero.
For the unicity of our solution, and the fact that a solution to the constraints of D'Hoker and Phong does not exist, we refer to a future paper \cite{DG}.

The constraints, and the well known one loop chiral measure, also determine
the modular forms $\Xi_6[\delta]$ of \cite{DPg2a},
\cite{DPg2b} uniquely, as we show in this paper. 
In particular, if one could prove a priori that the two loop chiral measure has the form indicated in section \ref{measures}, 
formula \eqref{guess} with $g=2$,
then we would have an easy derivation for the explicit formula of this measure.

In this paper we will use action of the modular group on modular forms as a powerful tool to solve the problem.
We took advantage of the theory of induced representations: the representation furnished by the space of forms is built up from the representation
given by a suitable subspace left invariant by a subgroup of the entire modular group.
This approach is similar to the method that Wigner used to classify the irreducible representations of the Poincar\'e group induced from the
representation of the little group. 
A systematic account of the representation of modular group on modular forms of
genus three and level two will also be given in \cite{DG}.

A delicate point is the assumption, made by D'Hoker and Phong and also in this paper,  that \eqref{ampiezgenh} is true also for three loops.
Moreover we will take contributions from even characteristics only. Even if there are many arguments leading to the conclusion that these last assumptions 
should not be valid for $g>2$ (see for example the discussion in \cite{CP}), we think that existence, uniqueness and simplicity of the solution are strong
arguments for the opposite conclusion, at least for $g=3$.
Also for $g=4$ it seems that the modified constraints have a solution,
we hope to report on this in the near future. 

\

The structure of the paper is the following.\\
In section \ref{section constraints} we recall some results on the measures for the bosonic string and the chiral superstring.
This leads us to consider the possibility that the chiral superstring measure might be obtained from
the bosonic string measure by multiplication by a modular form of weight $8$.

In section \ref{constraints} we formulate some constraints which this modular form should satisfy.
These constraints are very similar to the ones considered by D'Hoker and Phong in \cite{DP2}, the differences are discussed
in section \ref{differences}. It turns out that for genus two
we recover the chiral superstring measure as determined by D'Hoker and Phong \cite{DPg2a}, \cite{DPg2b}.
In genus three our constraints have a (unique) solution.

In section \ref{Siegel modular forms} we study the Siegel modular forms on $\Gamma_g(2)$ for $g=1,2$.
In particular, we show how our constraints lead to the functions $\Xi_6[\delta]$ in genus two found earlier by D'Hoker and Phong.
Our explicit formula for these functions in section \ref{g=2} is different from theirs and might be of independent interest.

In section \ref{g=3} we show that our constraints have a solution in genus three, and show that the corresponding cosmological
constant is zero.

In section \ref{conjecture} we briefly discuss some results of Igusa
which are related to a conjecture of D'Hoker and Phong.

In appendix A (section \ref{A}) we discuss characteristics and
symplectic geometry in a vector space over a field with two elements.
In appendix B (section \ref{appendix trans}) we recall some facts on the transformation theory of theta constants.
In appendix C (section \ref{res12}) we determine the restriction
of certain modular forms to `reducible' period matrices.

\

\section{Measures and Modular Forms}\label{section constraints}

\subsection{Basic definitions}\label{basic defs}
The Siegel upper half space of complex $g\times g$ symmetric matrices with positive definite imaginary part will be denoted by $\HH_g$. The action
of $\Gamma_g:=Sp(2g,\ZZ)$ on $\HH_g$ is denoted as usual by
$$
M\cdot \tau\,:=\,(A\tau+B)(C\tau+D)^{-1}\qquad
M:=\begin{pmatrix}A&B\\C&D\end{pmatrix}\in Sp(2g,\ZZ),\quad\tau\in\HH_g.
$$
A Siegel modular form $f$ of genus $g$ and weight $k$
on a subgroup $\Gamma \subset Sp(2g,\ZZ)$
is a holomorphic function on $\HH_g$ which satisfies
$$
f:\,\HH_g\longrightarrow\CC,\qquad f(M\cdot \tau)\,=\,\det(C\tau+D)^kf(\tau)
\qquad\forall M\in\Gamma,\quad\tau\in\HH_g
$$
(and in case $g=1$ one should also impose a growth condition on $f$).
The factor $\det(C\tau+D)$ satisfies a cocycle condition:
$$
\gamma(MN,\tau)\,=\,\gamma(M,N\cdot \tau)\gamma(N,\tau),\qquad
\mbox{where}\quad
\gamma(M,\tau)\,:=\,\det(C\tau+D).
$$
For a subgroup $\Gamma$ of $Sp(2g,\ZZ)$ which acts without fixed points on $\HH_g$
one can then define a linebundle, the Hodge bundle $\lambda$, on the quotient $\Gamma\backslash\HH_g$ as the quotient of the trivial bundle $\CC\times\HH_g$ on $\HH_g$
by the action of $\Gamma$ given by $\gamma$:
$$
\lambda:=\Gamma\backslash(\CC\times\HH_g)\longrightarrow \Gamma\backslash\HH_g,\qquad
M\cdot(t,\tau):=(\gamma(M,\tau)t,M\cdot \tau).
$$
Even in case $\Gamma$ has fixed points, we can use this action to define a sheaf, still called $\lambda$, on $\Gamma\backslash\HH_g$. The global sections of $\lambda$ correspond to the Siegel modular forms of
weight $1$; more generally, Siegel modular forms of weight $k$ correspond
to sections of $\lambda^{\otimes k}$.
For the definition of the well-known theta constants $\theta[\Delta](\tau)$ with even characteristics $\Delta$, which are modular forms of weight $1/2$ on a subgroup of $Sp(2g,\ZZ)$,
see section \ref{Heisenberg}.
Recall that there are $2^{g-1}(2^g+1)$ even characteristics.

\subsection{Measures}\label{measures}
We recall some results on the bosonic measure for $g\leq 3$,
the chiral superstring measure for $g\leq 2$ and on the proposal of
D'Hoker and Phong for the chiral superstring measure for $g=3$. In this section, $c_g$ and $c_g'$ are constants.

The genus one bosonic measure is
$$
{\rm d}\mu^{(1)}_B\,=\,
\frac{1}{(2\pi)^{12}\eta^{24}(\tau^{(1)})}
{\rm d}\tau^{(1)}.
$$
The genus one chiral superstring measure is (cf.\ e.g.\ \cite{DPg2b},
eq. (8.2))
$$
{\rm d}\mu[\Delta^{(1)}]\,=\,
\frac{\theta[\Delta^{(1)}]^4(\tau^{(1)})}{2^5\pi^4\eta^{12}(\tau^{(1)})}
{\rm d}\tau^{(1)},
$$
so that
$$
{\rm d}\mu[\Delta^{(1)}]\,=\,c_1'
\theta[\Delta^{(1)}]^4(\tau^{(1)})\eta^{12}(\tau^{(1)})\,{\rm d}\mu^{(1)}_B.
$$
Note that $\theta[\Delta^{(1)}]^4(\tau^{(1)})\eta^{12}(\tau^{(1)})$
is a modular form of weight $2+6=8$ on a subgroup of $SL(2,\ZZ)$.

The genus two bosonic measure is (\cite{BKMP}, \cite{M}):
$$
{\rm d}\mu^{(2)}_B\,=\,\frac{c_2}{\Psi_{10}(\tau^{(2)})}\prod_{i\leq j}\,{\rm d}\tau_{ij}
$$
where $\Psi_{10}$ is a modular form of weight $10$ on $Sp(4,\ZZ)$.
The genus two chiral superstring measure is (cf.\ \cite{DPg2a},
\cite{DPg2b}):
$$
{\rm d}\mu[\Delta^{(2)}]\,=\,
\frac{\theta[\Delta^{(2)}]^4(\tau^{(2)})\Xi_6[\Delta^{(2)}](\tau^{(2)})}
{16\pi^6\Psi_{10}(\tau^{(2)})}\prod_{i\leq j}\,{\rm d}\tau_{ij}
$$
so that
$$
{\rm d}\mu[\Delta^{(2)}]\,=\,c_2'
\theta[\Delta^{(2)}]^4(\tau^{(2)})\Xi_6[\Delta^{(2)}](\tau^{(2)})\,
{\rm d}\mu^{(2)}_B.
$$
Note that $\theta[\Delta^{(2)}]^4(\tau^{(2)})\Xi_6[\Delta^{(2)}](\tau^{(2)})$
is a modular form of weight $2+6=8$ on a subgroup of $Sp(4,\ZZ)$.

The genus three bosonic measure is (\cite{BKMP}, \cite{M}):
$$
{\rm d}\mu^{(3)}_B\,=\,\frac{c_3}{\Psi_{9}(\tau^{(3)})}
\prod_{i\leq j}\,{\rm d}\tau_{ij}
$$
where $\Psi_{9}^2(\tau^{(3)})$ is a Siegel modular form of weight $18$
for $Sp(6,\ZZ)$ (cf.\ \cite{Ic} for $\Psi_{9}$).

In \cite{DP2}, eqn.\ (1.1), D'Hoker and Phong propose that the genus three chiral superstring measure is of the form
$$
{\rm d}\mu[\Delta^{(3)}]\,=\,
\frac{\theta[\Delta^{(3)}]^4(\tau^{(3)})\Xi_6[\Delta^{(3)}](\tau^{(3)})}
{8\pi^4\Psi_{9}(\tau^{(3)})}\prod_{i\leq j}\,{\rm d}\tau_{ij},
$$
and they give three constraints on the functions
$\Xi_6[\Delta^{(3)}](\tau^{(3)})$. However, they do not succeed in finding
functions which satisfy all their constraints
(we will prove that there are indeed no such functions in \cite{DG}).

This leads us to weaken the proposal of \cite{DP2} and to search for
functions $\Xi_8[\Delta^{(3)}](\tau^{(3)})$, these should behave like the
products $\theta[\Delta^{(3)}]^4(\tau^{(3)})\Xi_6[\Delta^{(3)}]$ which  occur
in the numerator of ${\rm d}\mu[\Delta^{(3)}]$. So we assume that for $g=3$:
\begin{equation}\label{guess}
{\rm d}\mu[\Delta^{(g)}]\,=\,
c_g'\Xi_8[\Delta^{(g)}](\tau^{(g)}){\rm d}\mu^{(g)}_B
\end{equation}
where the functions $\Xi_8[\Delta^{(3)}](\tau^{(3)})$ satisfy three
constraints which are obtained from those imposed on the functions
$\theta[\Delta^{(3)}]^4(\tau^{(3)})\Xi_6[\Delta^{(3)}](\tau^{(3)})$
in \cite{DP2}. For example, in \cite{DP2} the function
$\Xi_6[\Delta^{(3)}](\tau^{(3)})$ is required to be
a Siegel modular form of weight $6$ on a subgroup
of $Sp(6,\ZZ)$, and we require that $\Xi_8[\Delta]$ is a
Siegel modular form of weight $2+6=8$ on a subgroup
of $Sp(6,\ZZ)$.
We give the three constraints on the $\Xi_8[\Delta^{(3)}]$ in section \ref{constraints}.
The main result of this paper is that functions
$\Xi_8[\Delta^{(3)}]$ which satisfy all three constraints
actually exist, the uniqueness of such functions will be shown in
\cite{DG}.

In general, one should keep in mind that since the canonical bundle
on the moduli space $M_g$ of genus $g$ curves is $13\lambda$, where
$\lambda$ is the Hodge bundle, the bosonic measure ${\rm d}\mu^{(g)}_B$ transforms as
a modular form of weight $-13$ (so the sections of $f{\rm d}\mu^{(g)}_B$ of the canonical bundle correspond to modular forms $f$ of weight $13$).
The chiral superstring measure
is known to transform as a section of $-5\lambda$ under the
action of a subgroup of $Sp(2g,\ZZ)$ (see \cite{D'Hoker:1989rk} and \cite{schwarz}, \cite{voronov} for the supersymmetric case).
Thus, taking only the transformation behaviour into account, it is not unreasonable to expect that equation (\ref{guess}) should hold for general $g$,
for some function
$\Xi_8[\Delta]$ which corresponds to a section of $8\lambda$.
As we just observed, this is proven to work for $g=1,2$ and for $g=3$
we can at least find unique $\Xi_8[\Delta^{(3)}]$ which satisfy
reasonable constraints.

\subsection{The modular forms $\Xi_8[\Delta]$}\label{constraints}
The discussion above thus leads us to search for functions
$$
\Xi_8[\Delta^{(g)}]\,:\,\HH_g\,\longrightarrow\,\CC,
\qquad\mbox{where}\quad
\Delta^{(g)}=[{}^{a_1\ldots a_g}_{b_1\ldots b_g}]
$$
is an even characteristic, that is $a_i,b_i\in\{0,1\}$ and
$\sum a_ib_i\equiv 0$ mod $2$. It is convenient to define
$\Xi_8[\Delta^{(g)}]=0$ in case $\Delta$ is an odd characteristic.

Actually, the $\Xi_8[\Delta^{(g)}]$'s should be defined on the subvariety $J_g\subset \HH_g$ of period matrices of Riemann surfaces
of genus $g$ (note that $\dim J_g=3g-3$ and $\dim \HH_g=g(g+1)/2$).
As we do not consider the cases $g>3$ in this paper we will write
$\HH_g$ instead of $J_g$.

In order to formulate constraints for these functions for all $g$, we
require that in case $g=1$ one has:
$$
\Xi_8[\Delta^{(1)}](\tau)\,:=\,
\theta[\Delta^{(1)}](\tau)^4\eta(\tau)^{12},
$$
see section \ref{measures}. Then, almost copying \cite{DP2},
we impose the constraints:

\

\noindent
(i) The functions $\Xi_8[\Delta^{(g)}]$ are holomorphic on $\HH_g$.

\noindent
(ii)
Under the action of $Sp(2g,\ZZ)$ on $\HH_g$, these functions should
transform as follows:
\begin{equation}\label{trasxi}
\Xi_8[M\cdot\Delta^{(g)}](M\cdot\tau)\,=\,
\det(C\tau+D)^8\Xi_8[\Delta^{(g)}](\tau),
\end{equation}
for all $M\in Sp(2g,\ZZ)$, here the action of $M$ on the characteristic $\Delta$ is given by
$$
\begin{pmatrix}A&B\\C&D\end{pmatrix}\cdot [{}^a_b]\,:=\,
[{}^c_d],\qquad
\left(\begin{array}{c}{}^tc\\{}^td\end{array}\right)\,=\,
\left(\begin{array}{cc} D&-C\\-B&A\end{array}\right)
\left(\begin{array}{c}{}^ta\\{}^tb\end{array}\right)
\,+\,
\left(\begin{array}{c}(C{}^tD)_0\\(A{}^tB)_0\end{array}\right)
\quad \mbox{mod}\;2
$$
where $N_0=(N_{11},\ldots,N_{gg})$ is the diagonal of the matrix $N$.

\noindent
(iii)
The restriction of these functions
to `reducible' period matrices is a product of the corresponding
functions in lower genus. More precisely, let
$$
\Delta_{k,g-k}\,:=\,\left\{\tau_{k,g-k}\,:=\,
\begin{pmatrix}\tau_k&0\\0&\tau_{g-k}\end{pmatrix}\,\in\HH_g\,:\,
\tau_k\in \HH_k,\;\tau_{g-k}\in\HH_{g-k}\,\right\}\;\cong\;
\HH_k\times\HH_{g-k}.
$$
Then we require that for all $k$, $0<k<g$,
$$
\Xi_8[{}^{a_1\ldots a_k\,a_{k+1}\ldots a_g}_{b_1\ldots b_k\,b_{k+1}\ldots b_g}](\tau_{k,g-k})\,=\,
\Xi_8[{}^{a_1\ldots a_k}_{b_1\ldots b_k}](\tau_k)
\Xi_8[{}^{a_{k+1}\ldots a_g}_{b_{k+1}\ldots b_g}](\tau_{g-k})
$$
for all even characteristics
$\Delta^{(g)}=[{}^{a_1\ldots a_g}_{b_1\ldots b_g}]$ and all
$\tau_{k,g-k}\in \Delta_{k,g-k}$.

\subsection{Remark: comparison with \cite{DP2}} \label{differences}
We compare these constraints with those of D'Hoker and Phong in
\cite{DP2} for the functions $\Xi_6[\Delta]$ on $\HH_3$.
The only essential difference is in constraint (ii). Note that
the products $\theta[\Delta]^4(\tau)\Xi_6[\Delta](\tau)$, with
$\Xi_6[\Delta]$ as in their constraint (ii) and $\tau\in\HH_3$,
transforms in the same way as our $\Xi_8[\Delta]$ but with
a factor $\epsilon(M,\Delta)^{4+4}$. However $\epsilon(M,\Delta)^8=1$,
so $\theta[\Delta]^4(\tau)\Xi_6[\Delta](\tau)$ transforms as $\Xi_8[\Delta]$.
Conversely, if each $\Xi_8[\Delta]$ were a product of $\theta[\Delta]^4$ and another function, these other functions would satisfy constraint (ii) of \cite{DP2}.

\subsection{Remark on condition (ii)}\label{remark on ii}
Let $\Gamma_g(2)$ be the (normal) subgroup  of
$Sp(2g,\ZZ)$ defined by:
$$
\Gamma_g(2)=\ker(Sp(2g,\ZZ)\,\longrightarrow\,Sp(2g,\FF_2))\,=\,
\{M\in Sp(2g,\ZZ):\;A\equiv D\equiv I,\;B\equiv C\equiv 0\;\mbox{mod}\;2\,\},
$$
where we write $\FF_2:=\ZZ/2\ZZ$ for the field with two elements.
For $M\in \Gamma_g(2)$ we have
$M\cdot[{}^a_b]=[{}^a_b]$ for all characteristics $[{}^a_b]$, hence
the $\Xi_8[\Delta^{(g)}]$ are modular forms of genus $g$ and weight $8$ on $\Gamma_g(2)$.

\subsection{Remark on condition (iii)} \label{arbitrary}
In \cite{DP2}, the third constraint is
imposed for an arbitrary separating degeneration. However,
any such degeneration is obtained from the one in condition (iii) by a symplectic transformation. Thus one has to consider the functions $\Xi_8[\Delta](N\cdot\tau_{k,g-k})$ for all $N\in Sp(2g,\ZZ)$. Constraint (ii) shows that this amounts to
considering $\Xi_8[N^{-1}\cdot \Delta](\tau_{k,g-k})$ (up to an easy factor)
and constraint (iii) determines this function.

\subsection{Reduction to the case $[\Delta^{(g)}]=[{}^0_0]$.}\label{O+}
The second constraint, in particular equation \ref{trasxi}, can be used to restrict the search for the
$2^{g-1}(2^g+1)$ functions $\Xi_8[\Delta^{(g)}]$ to that of a single one, for which we choose $\Xi_8[{}^0_0]$ with $[{}^0_0]=[{}^{0\ldots 0}_{0\ldots 0}]$. We work out the details of this reduction. In particular, we give the constraints which the function $\Xi_8[{}^0_0]$ should satisfy
and given this function we define functions $\Xi_8[\Delta^{(g)}]$,
for all even characteristics $\Delta^{(g)}$, which satisfy the constraints
from section \ref{constraints}.

Let $\Gamma_g(1,2)$ be the subgroup of $Sp(2g,\ZZ)$
which fixes the characteristic $[{}^0_0]:=[{}^{0\ldots0}_{0\ldots0}]$:
$$
\Gamma_g(1,2):=\{M\in \Gamma_g:\; M\cdot[{}^0_0]\equiv[{}^0_0]\;\mbox{mod}\,2\}
\,=\,
\{M\in \Gamma_g:\;{\rm diag}A{}^tB \equiv {\rm diag }C{}^tD\equiv \,0\;{\rm mod}\,2\,\}.
$$
For $M\in \Gamma_g(1,2)$ we required that
$\Xi_8[{}^0_0](M\cdot\tau)=(C\tau+D)^8\Xi_8[{}^0_0](\tau)$, that is,
$\Xi_8[{}^0_0]$ is a modular form on $\Gamma_g(1,2)$ of weight $8$.

Given such a modular form $\Xi_8[{}^0_0]$ on $\Gamma_g(1,2)$
we now define, for each even characteristic $\Delta$
a function $\Xi_8[\Delta]$ in such a way that equation \ref{trasxi} holds.
As the group $Sp(2g,\ZZ)$ acts transitively on the even characteristics,
for any even characteristic $[\Delta^{(g)}]$ there is an
$M\in Sp(2g,\ZZ)$ with $M\cdot [{}^0_0]=[\Delta^{(g)}]$ mod $2$
and then we define, with $\gamma$ as in section \ref{basic defs},
\begin{equation}\label{defxidelta}
\Xi_8[\Delta^{(g)}](\tau):=
\gamma(M,M^{-1}\cdot\tau)^8\Xi_8[{}^0_0](M^{-1}\cdot \tau).
\end{equation}

It is easy to check that the definition of $\Xi_8[\Delta^{(g)}]$ does not
depend on the choice of $M$: if also $N\cdot [{}^0_0]=[\Delta^{(g)}]$ mod $2$, then $N^{-1}M$ fixes $[{}^0_0]$ so $N^{-1}M\in\Gamma_g(1,2)$.
To verify that
$$
\gamma(M,M^{-1}\cdot\tau)^8\Xi_8[{}^0_0](M^{-1}\cdot \tau)
\,\stackrel{?}{=}
\gamma(N,N^{-1}\cdot\tau)^8\Xi_8[{}^0_0](N^{-1}\cdot \tau)
$$
we let $\tau=M\tau'$, so we must verify that
$$
\gamma(M,\tau')^8\Xi_8[{}^0_0]( \tau')\,\stackrel{?}{=}
\gamma(N,N^{-1}M\cdot\tau')^8\Xi_8[{}^0_0](N^{-1}M\cdot \tau').
$$
As $N^{-1}M\in \Gamma_g(1,2)$ and $\gamma$ satisfies the cocycle condition, we get
{\renewcommand{\arraystretch}{1.5}
$$
\begin{array}{rcl}
\gamma(N,N^{-1}M\cdot\tau')^8\Xi_8[{}^0_0](N^{-1}M\cdot \tau')&=&
\gamma(N,N^{-1}M\cdot\tau')^8\gamma(N^{-1}M,\tau')^8\Xi_8[{}^0_0](\tau')\\
&=&\gamma(M,\tau')^8\Xi_8[{}^0_0](\tau'),
\end{array}
$$
}
which verifies the desired identity. Finally we show that the functions
$ \Xi_8[\Delta^{(g)}]$ satisfy constraint (ii) of section \ref{constraints}.
So with $M,\Delta^{(g)}$ as above, we must verify that for all $N\in Sp(2g,\ZZ)$ we have
$$
\Xi_8[N\cdot\Delta](N\cdot\tau)\,\stackrel{?}{=}\,
\gamma(N,\tau)^8\Xi_8[\Delta](\tau).
$$
As $N\cdot\Delta=NM\cdot[{}^0_0]$, we have:
{\renewcommand{\arraystretch}{1.5}
$$
\begin{array}{rcl}
\Xi_8[N\cdot\Delta](N\cdot\tau)&=&
\gamma(NM,(NM)^{-1}N\cdot\tau)^8\Xi_8[{}^0_0]((NM)^{-1}N\cdot\tau)\\
&=&\gamma(NM,M^{-1}\cdot\tau)^8\Xi_8[{}^0_0](M^{-1}\tau)\\
&=&\gamma(N,\tau)^8\gamma(M,M^{-1}\cdot\tau)^8
\Xi_8[{}^0_0](M^{-1}\cdot\tau)\\
&=&\gamma(N,\tau)^8\Xi_8[\Delta](\tau),
\end{array}
$$
}
where we used the cocycle relation. Thus the second constraint is verified
if $\Xi_8[{}^0_0]$ satisfies the constraint (ii$_0$) below and if
the $\Xi_8[\Delta^{(g)}]$ are defined as in equation \ref{defxidelta}.

\

\noindent
(ii$_0$) The function $\Xi_8[{}^0_0]$
is a modular form $\Xi_8$ of weight $8$ on $\Gamma_g(1,2)$.

\

Next we show that, in case $g\leq3$,
constraint (iii) follows from the constraints:

\

\noindent
(iii$_0$)(1) For all $k$, $0< k <g$, and all $\tau_{k,g-k}\in \Delta_{k,g-k}$ we have
$$
\Xi_8[{}^0_0](\tau_{k,g-k})\,=\,
\Xi_8[{}^0_0](\tau_{k})\Xi_8[{}^0_0](\tau_{g-k})
$$
(iii$_0$)(2) If $\Delta^{(g)}=[{}^{ab\ldots}_{cd\ldots}]$ with $ac=1$ then
$\Xi_8[\Delta^{(g)}](\tau_{1,g-1})=0$.

\

Obviously (iii) implies (iii$_0$)(1,2).
We will only show how to use (iii$_0$)(1).
Let
$$
\Delta^{(g)}=[{}^{a_1\ldots a_g}_{b_1\ldots b_g}],\qquad
\Delta^{(k)}:=[{}^{a_1\ldots a_k}_{b_1\ldots b_k}],\quad
\Delta^{(g-k)}:=[{}^{a_{k+1}\ldots a_g}_{b_{k+1}\ldots b_g}].
$$
and assume that $\Delta^{(k)}$ is even,
then also $\Delta^{(g-k)}$ is even. Thus there are symplectic matrices $M_1\in Sp(2k,\ZZ)$ and
$M_2\in Sp(2(g-k),\ZZ)$ such that $M_1\cdot[{}^0_0]=[\Delta^{(k)}]$
and $M_2\cdot[{}^0_0]=[\Delta^{(g-k)}]$.
Hence the matrix
$M\in Sp(2g,\ZZ)$ obtained from $M_1,M_2$ in the obvious way has
the properties: $M\cdot(\Delta_{k,g-k})=\Delta_{k,g-k}$
and $\Delta^{(g)}=M\cdot [{}^0_0]$.
As $M,\tau_{k,g-k}$ are made up of $k\times k$ and $(g-k)\times(g-k)$ blocks
one has
$$
\gamma(M,M^{-1}\cdot\tau_{k,g-k})=\gamma(M_1,M_1^{-1}\cdot\tau_k)
\gamma(M_2,M_2^{-1}\cdot\tau_{g-k}),\qquad \Delta^{(g)}=M\cdot [{}^0_0].
$$
Moreover, $M^{-1}\cdot\tau_{k,g-k}$ is the matrix in $\Delta_{k,g-k}$ with
blocks $M_1^{-1}\cdot\tau_k$ and $M_2^{-1}\cdot\tau_{g-k}$. Thus if (iii$_0$)(1)
is satisfied we have:
$$
\Xi_8[{}^0_0](M^{-1}\cdot\tau_{k,g-k})=
\Xi_8[{}^0_0](M_1^{-1}\cdot\tau_k)\Xi_8[{}^0_0](M_2^{-1}\cdot\tau_{g-k}).
$$

Then we have:
{\renewcommand{\arraystretch}{1.5}
$$
\begin{array}{rcl}
\Xi_8[\Delta^{(g)}](\tau_{k,g-k})&=&\gamma(M,M^{-1}\cdot\tau)^8
\Xi_8[{}^0_0](M^{-1}\cdot\tau_{k,g-k})\\
&=&
\gamma(M_1,M_1^{-1}\cdot\tau_k)^8
\gamma(M_2,M_2^{-1}\cdot\tau_{g-k})^8
\Xi_8[{}^0_0](M_1^{-1}\cdot\tau_k)\Xi_8[{}^0_0](M_2^{-1}\cdot\tau_{g-k})\\
&=&\Xi_8[\Delta^{(k)}](\tau_k)\Xi_8[\Delta^{(g-k)}](\tau_{g-k}),
\end{array}
$$
}
so for these $\Delta^{(g)}$ the functions $\Xi_8[\Delta^{(g)}]$
satisfy (iii).

\section{Siegel modular forms}
\label{Siegel modular forms}

\subsection{Theta constants}\label{Heisenberg}
Modular forms of even weight on $\Gamma_g(2)$
can be obtained as products of the even theta constants
$$
\theta[{}^a_b](\tau)\,:=\,\sum_{m\in\ZZ^g}
\,e^{\pi i({}^t(m+a/2)\tau(m+a/2)+{}^t(m+a/2)b)}
$$
with $a=(a_1,\ldots,a_g),\;b=(b_1,\ldots,b_g)$, $a_i,b_i\in\{0,1\}$
and $\sum a_ib_i\equiv 0$ mod $2$ (note that we write $a,b$ as row vectors
in $[{}^a_b]$ but as column vectors in $m+a/2,{}^t(m+a/2)b$).
These theta constants are modular forms, of weight $1/2$, for a subgroup
of $Sp(2g,\ZZ)$.

It is convenient to
define the $2^g$ theta constants:
$$
\Theta[\sigma](\tau)\,:=\, \theta[{}^\sigma_{0}](2\tau),\qquad
[\sigma]=[\sigma_1\;\sigma_2\;\ldots\;\sigma_g],\;\sigma_i\in\{0,1\},\;
\tau\in\HH_g.
$$
The $\Theta[\sigma]$ have the advantage that they are algebraically independent for $g\leq 2$ and there is a unique relation of degree 16 for $g=3$, whereas there are many algebraic relations
between the $\theta[\Delta]$'s, for example Jacobi's relation in $g=1$.

\subsection{A classical formula}\label{classical}
A classical formula for theta functions shows that any $\theta[\Delta]^2$
is a linear combination of products of two $\Theta[\sigma]$'s.
Note that there are $2^g$ functions $\Theta[\sigma]$ and thus there are
$(2^g+1)2^g/2=2^{g-1}(2^g+1)$ products $\Theta[\sigma]\Theta[\sigma']$.
This is also the number of even characteristics, and the products
$\Theta[\sigma]\Theta[\sigma']$ span the same space (of modular forms of weight 1) as the $\theta[\Delta]^2$'s.

The classical formula used here is (cf.\ \cite{Igusa}, IV.1, Theorem 2):
$$
\theta[{}^\epsilon_{\epsilon'}]^2\,=\,\sum_{\sigma}
(-1)^{\sigma\epsilon'}\Theta[\sigma]\Theta[\sigma+\epsilon]
$$
where we sum over the $2^g$ vectors $\sigma$ and $[{}^\epsilon_{\epsilon'}]$
is an even characteristic, so $\epsilon\epsilon'\equiv 0$ mod $2$.
These formulae are easily inverted to give:
$$
\Theta[\sigma]\Theta[\sigma+\epsilon]\,=\,\mbox{$\frac{1}{2^g}$}\sum_{\epsilon'}
(-1)^{\sigma\epsilon'}\theta[{}^\epsilon_{\epsilon'}]^2.
$$

\subsubsection{Example}
In case $g=1$ one has
$$
\theta[{}^0_0]^2=\Theta[0]^2+\Theta[1]^2,\qquad
\theta[{}^0_1]^2=\Theta[0]^2-\Theta[1]^2,\qquad
\theta[{}^1_0]^2=2\Theta[0]\Theta[1],
$$
or, equivalently,
$$
\Theta[0]^2=(\theta[{}^0_0]^2+\theta[{}^0_1]^2)/2,\qquad
\Theta[1]^2=(\theta[{}^0_0]^2-\theta[{}^0_1]^2)/2,\qquad
\Theta[0]\Theta[1]=\theta[{}^1_0]^2/2.
$$
Note that upon substituting the first three relations in Jacobi's relation
$\theta[{}^0_0]^4=\theta[{}^0_1]^4+\theta[{}^1_0]^4$ one obtains a trivial
identity in the $\Theta[\sigma]$'s.

\subsection{The case $g=1$}\label{g=1}\label{dec g=1}
In the genus one case, the modular forms $\Xi_8[\Delta]$ are given by $\Xi_8[\Delta]=\theta[\Delta]^4\eta^{12}$.
A classical formula for the Dedekind $\eta$ function is:
$\eta^3=\theta[{}^0_0]\theta[{}^0_1]\theta[{}^1_0]$, so
{\renewcommand{\arraystretch}{1.5}
$$
\begin{array}{rcl}
\eta^{12}&=&\theta[{}^0_0]^4\theta[{}^0_1]^4\theta[{}^1_0]^4\\
&=&(\Theta[0]^2+\Theta[1]^2)^2
(\Theta[0]^2-\Theta[1]^2)^2
(2\Theta[0]\Theta[1])^2.
\end{array}
$$
}
Another useful formula for $\eta^{12}$ is closely related to Jacobi's relation:
$$
3\eta^{12}=\theta[{}^0_0]^{12}-\theta[{}^0_1]^{12}-\theta[{}^1_0]^{12},
$$
it suffices to use the classical formula's, expressing the $\theta[{}^a_b]$'s
in the $\Theta[\sigma]$'s, to verify the identity.

The function $\Xi_8[{}^0_0]=\theta[{}^0_0]^4\eta^{12}$ is a modular form
on $\Gamma_1(1,2)$ of weight eight. Another modular form of the same type is
$\theta[{}^0_0]^4f_{21}(\tau)$ with
$$
f_{21}\,:=\,2\theta[{}^0_0]^{12}+\theta[{}^0_1]^{12}+\theta[{}^1_0]^{12}.
$$
Below we write some other modular forms, which we will need later, in terms of $f_{21},\eta^{12}$:
{\renewcommand{\arraystretch}{1.5}
$$
\begin{array}{rcl}
\theta^{12}[{}^0_0]&=&\mbox{$\frac{1}{3}$}f_{21}+\eta^{12},\\
\theta[{}^0_0]^4( \theta[{}^0_0]^8+\theta[{}^0_1]^8+\theta[{}^1_0]^8)
&=&\theta[{}^0_0]^{12}+\theta[{}^0_0]^4
(\theta[{}^0_0]^8-2\theta[{}^0_1]^4\theta[{}^1_0]^4)\\
&=&(\mbox{$\frac{1}{3}$}f_{21}+\eta^{12})+(\mbox{$\frac{1}{3}$}f_{21}+
\eta^{12})-2\eta^{12}\\
&=&\mbox{$\frac{2}{3}$}f_{21},\\
\theta[{}^0_0]^{12}+\theta[{}^0_1]^{12}+\theta[{}^1_0]^{12}&=&
f_{21}-\theta[{}^0_0]^{12}\\
&=&\mbox{$\frac{2}{3}$}f_{21}-\eta^{12},\\
\theta[{}^0_0]^4\theta[{}^0_1]^8+
\theta[{}^0_0]^4\theta[{}^1_0]^8
&=&
\theta[{}^0_0]^4(\theta[{}^0_0]^8+\theta[{}^0_1]^8+\theta[{}^1_0]^8)
-\theta[{}^0_0]^{12}\\
&=&
\mbox{$\frac{2}{3}$}f_{21}-(\mbox{$\frac{1}{3}$}f_{21}+\eta^{12})\\
&=&\mbox{$\frac{1}{3}$}f_{21}-\eta^{12}.
\end{array}
$$
}

\subsection{The case $g=2$}\label{g=2}
In case $g=2$, we define three holomorphic functions on $\HH_2$:
$$
f_1:=\theta[{}^{00}_{00}]^{12},\qquad
f_2:=\sum_\delta\theta[\delta]^{12},\qquad
f_3:=\theta[{}^{00}_{00}]^4\sum_\delta\theta[\delta]^{8},
$$
where we sum over the $10$ even characteristics $\delta$ in genus $2$.
The functions $\theta[{}^{00}_{00}]^4f_i$, $i=1,2,3$, are modular forms
of weight $8$ for $\Gamma_2(1,2)$, see Appendix \ref{applications}.

The function $\sum_\delta \theta[\delta]^{16}$
is a modular form on $Sp(4,\ZZ)$, and hence is a modular form
of weight $8$ for $\Gamma_2(1,2)$, but we do not need it. In \cite{DG}
we will show that the three $\Theta[{}^{00}_{00}]^4f_i$ and
$\sum_\delta \theta[\delta]^{16}$
are a basis of the modular forms
of weight $8$ on $\Gamma_2(1,2)$.

The third constraint on the function $\Xi_8[{}^{00}_{00}]$ is:
$$
\Xi_8[{}^{00}_{00}](\tau_{1,1})\,=\,(\theta[{}^0_0]^4\eta^{12})(\tau_1)
(\theta[{}^0_0]^4\eta^{12})(\tau_1')
$$
where $\tau_{1,1}=diag(\tau_1,\tau_1')$ and $\tau_1,\tau_1'\in\HH_1$.
We try to determine $a_i\in\CC$ such that
$\theta[{}^{00}_{00}]^4\sum_ia_if_i$ factors in this way for such period matrices. Note that
$$
\theta[{}^{ab}_{cd}](\tau_{1,1})\,=\,
\theta[{}^{a}_{c}](\tau_{1})\theta[{}^{b}_{d}](\tau_{1}'),
$$
in particular $\theta[{}^{ab}_{cd}](\tau_{1,1})=0$ if $ac=1$.
As $\theta[{}^{00}_{00}](\tau_{1,1})$ produces $\theta[{}^0_0]^4(\tau_1)
\theta[{}^0_0]^4(\tau_1')$, it remains to find $a_i$ such that
$$
\eta^{12}(\tau_1)\eta^{12}(\tau_1')\,=\,(a_1f_1+a_2f_2+a_3f_3)(\tau_{1,1}).
$$
Using the results from \ref{dec g=1}, the restrictions of the $f_i$ are:
{\renewcommand{\arraystretch}{1.5}
$$
\begin{array}{rcl}
\theta[{}^{00}_{00}]^{12}(\tau_{1,1})&=&
\theta[{}^0_0]^{12}(\tau_{1})
\theta[{}^{0}_{0}]^{12}(\tau_{1}')\\
&=&(\mbox{$\frac{1}{3}$}f_{21}+\eta^{12})(\tau_1)
(\mbox{$\frac{1}{3}$}f_{21}+\eta^{12})(\tau_1'),
\\&&\\
(\sum_\delta\theta[\delta]^{12})(\tau_{1,1})&=&
(\theta[{}^0_0]^{12}+\theta[{}^0_1]^{12}+\theta[{}^1_0]^{12})(\tau_1)
(\theta[{}^0_0]^{12}+\theta[{}^0_1]^{12}+\theta[{}^1_0]^{12})(\tau_1')\\
&=&(\mbox{$\frac{2}{3}$}f_{21}-\eta^{12})(\tau_1)
(\mbox{$\frac{2}{3}$}f_{21}-\eta^{12})(\tau_1'),
\\&&\\
\Bigl(\theta[{}^{00}_{00}]^4\sum_\delta\theta[\delta]^8\Bigr)(\tau_{1,1})&=&
\Bigl(\theta[{}^0_0]^4
(\theta[{}^0_0]^8+\theta[{}^0_1]^8+\theta[{}^1_0]^8\Bigr)(\tau_1)
\Bigl(\theta[{}^{0}_{0}]^4
(\theta[{}^0_0]^8+\theta[{}^0_1]^8+\theta[{}^1_0]^8)\Bigr)(\tau_1')\\
&=&\mbox{$\frac{2}{3}$}f_{21}(\tau_1)
\mbox{$\frac{2}{3}$}f_{21}(\tau_1').
\end{array}
$$
} 
Next we require that the term $f_{21}(\tau_1)$ disappears in the linear
combination ($\sum a_if_i)(\tau_{1,1})$, that gives
$$
\Bigl(a_1(\mbox{$\frac{1}{3}$}f_{21}+\eta^{12})+
2a_2(\mbox{$\frac{2}{3}$}f_{21}-\eta^{12})+
2a_3\mbox{$\frac{2}{3}$}f_{21}\Bigr)(\tau_1')\,=\,0
$$
for all $\tau_1'\in \HH_1$. This gives two linear equations for the $a_i$ which have a unique solution, up to scalar multiple:
$$
a_1+4a_2+4a_3=0,\qquad a_1-2a_2=0,\qquad\mbox{hence}\quad
(a_1,a_2,a_3)=\lambda(-4,-2,3).
$$
A computation shows that $(4f_1+2f_2-3f_3)(\tau_{1,2})=6\eta^{12}(\tau_1)\eta^{12}(\tau_1')$.
Thus we conclude that
$$
\Xi_8[{}^{00}_{00}]\,:=\,\theta[{}^{00}_{00}]^4\bigl(
4\theta[{}^{00}_{00}]^{12}
+2\sum_\delta\theta[\delta]^{12}
-3\theta[{}^{00}_{00}]^4\sum_\delta\theta[\delta]^{8}\bigr)/6
$$
satisfies the constraints. We will show in \cite{DG} that this is the unique
modular form on $\Gamma_2(1,2)$ satisfying the constraints.

As $\theta[{}^{00}_{00}]^4\Xi_6[{}^{00}_{00}]$ satisfies the same constraints
(with $\Xi_6[{}^{00}_{00}]$ the modular form determined by D'Hoker and Phong
in \cite{DPg2a}, \cite{DPg2b}) we obtain from the uniqueness (or from a direct computation using the methods of \cite{CP}) that
$$
\Xi_6[{}^{00}_{00}]\,=\,
\Bigl(
4\theta[{}^{00}_{00}]^{12}
+2\sum_\delta\theta[\delta]^{12}
-3\theta[{}^{00}_{00}]^4\sum_\delta\theta[\delta]^{8}\Bigr)/6.
$$
Another formula for this function is:
$$
\Xi_6[{}^{00}_{00}]\,=\,-
(\theta[{}^{00}_{11}]\theta[{}^{01}_{00}]\theta[{}^{10}_{01}])^4 -
(\theta[{}^{00}_{01}]\theta[{}^{01}_{10}]\theta[{}^{11}_{00}])^4 -
(\theta[{}^{00}_{10}]\theta[{}^{10}_{00}]\theta[{}^{11}_{11}])^4,
$$
which is the one found by D'Hoker and Phong in \cite{DPg2b}, to check the
equality between the two expressions for $\Xi_6[{}^{00}_{00}]$
one can use the classical theta formula.

\section{The genus three case}\label{g=3}

\subsection{Modular forms in genus three}
\label{geo g=3}\label{weight 4} \label{weight 8}
In case $g=3$, the $8$ $\Theta[\sigma]$'s define a holomorphic map
$$
\HH_3\,\longrightarrow\, \PP^7,\qquad \tau\longmapsto
(\Theta[000](\tau):\ldots:\Theta[111](\tau)).
$$
The closure of the image of this map is a $6$-dimensional projective variety which is defined by a homogeneous polynomial $F_{16}$
(in eight variables) of degree $16$.
In particular, the holomorphic function
$\tau\mapsto F_{16}(\ldots,\Theta[\sigma](\tau),\ldots)$ is identically zero on $\HH_3$.

To write down $F_{16}$
we recall the following relation, which holds for all $\tau\in\HH_3$:
$$
r_1-r_2=r_3,\qquad {\rm with}\quad
r_1=\prod_{a,b\in\FF_2}\theta[{}^{000}_{0ab}](\tau),\quad
r_2=\prod_{a,b\in\FF_2}\theta[{}^{000}_{1ab}](\tau),\quad
r_3=\prod_{a,b\in\FF_2}\theta[{}^{100}_{0ab}](\tau).
$$
{F}rom this we deduce that $2r_1r_2=r_1^2+r_2^2-r_3^2$ and thus
$$
r_1^4+r_2^4+r_3^4-2(r_1^2r_2^2+r_1^2r_3^2+r_2^2r_3^2)
$$
is zero, as function of $\tau$, on $\HH_3$.
Let $F_{16}$ be the homogeneous polynomial,
of degree $16$ in the $\Theta[\sigma]$'s, obtained (using the
classical theta formula \ref{classical}) from this polynomial (of degree $8$) in the $\theta[\Delta]^2$.
In \cite{vGvdG} it is shown that $F_{16}$
is not zero as a polynomial in the eight $\Theta[\sigma]$.
Thus the polynomial $F_{16}$
defines the image of $\HH_3\rightarrow \PP^7$.

A computer computation, using once again the classical formula,
shows that
$F_{16}$ coincides, up to a scalar multiple, with the degree 16
polynomial in the $\Theta[\sigma]$ obtained from
$$
8\sum_\Delta\theta[\Delta]^{16}\;-\,
\Bigl(\sum_\Delta \,\theta[\Delta]^{8}\Bigr)^2
$$
by the classical theta formulas.

\subsection{The functions $F_i$}\label{Fi}
In analogy with the genus two case, we now want to find
a modular form $\Xi_8[{}^{000}_{000}]$ of weight $8$ on
$\Gamma_3(1,2)$
which restricts to the `diagonal' $\Delta_{1,2}$ as
$$
\Xi_8[{}^{000}_{000}](\tau_{1,2})\,=\,
\Xi_8[{}^0_0](\tau_1)\Xi_8[{}^{00}_{00}](\tau_2)\,=\,
\Bigl(\theta[{}^0_0]^4\eta^{12}\Bigr)(\tau_1)
\Bigl(\theta[{}^{00}_{00}]^4\Xi_6[{}^{00}_{00}]\Bigr)(\tau_2)
$$
where $\tau_{1,2}\in\HH_3$ is the block diagonal matrix with entries $\tau_1\in\HH_1$ and $\tau_2\in\HH_2$.
An obvious generalization of the functions $f_i$ which we considered
earlier in section \ref{g=2} are:
$$
F_1:=\,\theta[{}^{000}_{000}]^{12},\qquad
F_2:=\,\sum_\Delta \,\theta[\Delta]^{12},\qquad
F_3:=\,\theta[{}^{000}_{000}]^4\sum_\Delta\theta[\Delta]^8,
$$
where the sum is over the $36$ even characteristics $\Delta$ in genus
three. The functions $\theta[{}^{000}_{000}]^4F_i$
are modular forms of weight $8$ on $\Gamma_3(1,2)$,
see Appendix \ref{applications}.
However, there is no linear combination of these three functions
which has the desired restriction. Therefore we introduce another
modular form $G[{}^{000}_{000}]$ of weight $8$ on $\Gamma_3(1,2)$
in the next section.

\subsection{The modular forms $G[\Delta]$.}\label{Gdelta}
For each even characteristic $\Delta$ in $g=3$ we define
a modular form $G[\Delta]$ of weight $8$ on $\Gamma_3(2)$.
For a brief introduction to characteristics, quadrics and isotropic subspaces see Appendix \ref{quasymp}.

An even characteristic $\Delta$ corresponds to a quadratic form
$$
q_\Delta\,:\,V=\FF_2^6\,\longrightarrow\,\FF_2
$$
which satisfies $q_\Delta(v+w)=q_\Delta(v)+q_\Delta(w)+E(v,w)$
where
$E(v,w):=\sum_{i=1}^3(v_iw_{3+i}+v_{3+i}w_{i})$.
If $\Delta=[{}^{abc}_{def}]$ then:
$$
q_\Delta(v)=v_1v_4+v_2v_5+v_3v_6+av_1+bv_2+cv_3+dv_4+ev_5+fv_6,
$$
where $v=(v_1,\ldots,v_6)\in V$,
we will also write $v=({}^{v_1v_2v_3}_{v_4v_5v_6})$.
Let $Q_\Delta=\{v\in V:\,q_\Delta(v)=0\}$ be the corresponding
quadric in $V$.

A Lagrangian (i.e.\ maximally isotropic subspace) $L\subset V$ is
a subspace of $V$ such that $E(v,w)=0$ for
for all $v,w\in L$ and such that $\dim L=3$. For example,
the eight elements $({}^{abc}_{000})\in V$ with $a,b,c\in \FF_2$
form a Lagrangian subspace $L_0$ in $V$.

For such a subspace $L$ we define
a modular form on a subgroup of $Sp(6,\ZZ)$:
$$
P_L\,:=\,\prod_{Q\supset L}\theta[\Delta_Q]^2
$$
here the product is over the even quadrics which contain $L$
(there are eight such quadrics for each $L$)
and $\Delta_Q$ is the even characteristic corresponding to $Q$.
In case $L=L_0$ with
$$
L_0:=\{(v_1,\ldots,v_6)\in V:\,v_4=v_5=v_6=0\,\},\qquad
P_{L_0}=(r_1r_2)^2=\prod_{a,b,c\in\FF_2}\theta[{}^{000}_{abc}]^2
$$
with $r_1,r_2$ as in section \ref{weight 4}.
The action of $Sp(6,\ZZ)$ on $V=\ZZ^6/2\ZZ^6$ permutes the Lagrangian subspaces 
$L$, the subgroup $\Gamma_3(2)$ acts trivially on $V$. 
Similarly, the $P_L$ are permuted by the action of $Sp(6,\ZZ)$, see Appendix
\ref{trans PL}, and as $\Gamma_3(2)$ fixes all $L$'s, the $P_L$ are modular
forms on $\Gamma_3(2)$ of weight $8$.

For an even characteristic $\Delta$, the quadric $Q_\Delta$ contains
$30$ Lagrangian subspaces.
The sum of the $30$ $P_L$'s, with $L$ a Lagrangian subspace of $Q_\Delta$, is a modular form $G[\Delta]$ of weight $8$ on $\Gamma_3(2)$:
$$
G[\Delta]\,:=\,\sum_{L\subset Q_\Delta}\,P_L\,=\,
\,\sum_{L\subset Q_\Delta}\,
\prod_{Q'\supset L}\theta[\Delta_{Q'}]^2.
$$
Note that $\theta[\Delta]^2$ is one of the factors in each of the 30 products.
As the $P_L$ are permuted by the action of $Sp(6,\ZZ)$, also
the $G[\Delta]$ are permuted:
$$
G[M\cdot\Delta](M\cdot \tau)\,=\,\det(C\tau+D)^8 G[\Delta](\tau).
$$
As $\Gamma_3(1,2)$ fixes the characteristic $[^{000}_{000}]$,
the function $G[{}^{000}_{000}]$ is a modular form on $\Gamma_3(1,2)$.

\subsection{The restriction}
Now we try to find a linear combination of the functions $\theta[{}^{000}_{000}]^4F_i$, $i=1,2,3$
and $G[{}^{000}_{000}]$ which satisfies the third constraint:
$$
\Bigl(\theta[{}^0_0]^4\eta^{12}\Bigr)(\tau_1)
\Bigl(\theta[{}^{00}_{00}]^4\Xi_6[{}^{00}_{00}]\Bigr)(\tau_2)\,=\,
\Bigl(
\theta[{}^{000}_{000}]^4(b_1F_1+b_2F_2+b_3F_3)+b_4G[{}^{000}_{000}]
\Bigr)(\tau_{1,2}).
$$
It is easy to see that the theta constants satisfy:
$$
\theta[{}^{abc}_{def}](\tau_{1,2})\,=\,
\theta[{}^{a}_{d}](\tau_1)\theta[{}^{bc}_{ef}](\tau_2),
$$
in particular $\theta[{}^{abc}_{def}]\mapsto0$ if $ad=1$.
Thus $6$ of the $36$ even theta constants map to zero, the other
$30=3\cdot 10$ are uniquely decomposed in the product of two even theta constants for $g=1$ and $g=2$ respectively.
Using the results from \ref{dec g=1}, the functions $F_i(\tau_{1,2})$ are then easy to describe, the function $G[{}^{000}_{000}](\tau_{1,2})$ is determined in Appendix \ref{resg00}. The restrictions to 
$\Delta_{1,2}\cong \HH_1\times\HH_2$ are:
{\renewcommand{\arraystretch}{1.5}
$$
\begin{array}{rl}
(\theta[{}^{000}_{000}]^{12})_{|\Delta_{1,2}}=&
\theta[{}^0_0]^{12}
\theta[{}^{00}_{00}]^{12}\\
=&(\mbox{$\frac{1}{3}$}f_{21}+\eta^{12})\theta[{}^{00}_{00}]^{12},
\\&\\
(\sum_\Delta\theta[\Delta]^{12})_{|\Delta_{1,2}}=&
(\theta[{}^0_0]^{12}+\theta[{}^0_1]^{12}+\theta[{}^1_0]^{12})
(\sum_\delta \theta[\delta]^{12})\\
=&(\mbox{$\frac{2}{3}$}f_{21}-\eta^{12})\sum_\delta \theta[\delta]^{12},
\\&\\
(\theta[{}^{000}_{000}]^4\sum_\Delta \!\theta[\Delta]^8)_{|\Delta_{1,2}}
=&\theta[{}^0_0]^4
(\theta[{}^0_0]^8+\theta[{}^0_1]^8+\theta[{}^1_0]^8)
\theta[{}^{00}_{00}]^4
(\sum_\delta \theta[\delta]^{8})\\
=&\mbox{$\frac{2}{3}$}f_{21}
\bigl(\theta[{}^{00}_{00}]^4(\sum_\delta \theta[\delta]^{8})\bigr),
\\&\\
G[{}^{000}_{000}]_{|\Delta_{1,2}}=&
\bigl(\theta[{}^0_0]^4(\mbox{$\frac{1}{3}$}f_{21}-\eta^{12})\bigr)
\bigl(\theta[{}^{00}_{00}]^4
(\mbox{$\frac{1}{3}$}\theta[{}^{00}_{00}]^{12}
+\mbox{$\frac{2}{3}$}\sum_\delta\theta[\delta]^{12}
-\mbox{$\frac{1}{2}$}\theta[{}^{00}_{00}]^{4}\sum_\delta\theta[\delta]^{8})\bigr).\!\!
\\&
\end{array}
$$
} 

Thus we found the restriction of $
\theta[{}^{000}_{000}]^4(b_1F_1+b_2F_2+b_3F_3)+b_4G[{}^{000}_{000}]
$ to $\Delta_{1,2}$,
note that the function $\theta[{}^{000}_{000}]^4$
in front of the $F_i$ gives the function $\theta[{}^0_0]^4(\tau_1)\theta[{}^{00}_{00}]^4(\tau_2)$.
In particular, the restriction has a factor $\theta[{}^0_0]^4\theta[{}^{00}_{00}]^4$. In order that this restriction
is a multiple of $\theta[{}^0_0]^4\eta^{12}$ we need that the term $f_{21}$ disappears, which leads to the equation
$$
b_1\theta[{}^{00}_{00}]^{12}+2b_2\!\sum_\delta \theta[\delta]^{12}
+2b_3\theta[{}^{00}_{00}]^4(\sum_\delta \theta[\delta]^{8})
+b_4(\mbox{$\frac{1}{3}$}\theta[{}^{00}_{00}]^{12}
+\mbox{$\frac{2}{3}$}\sum_\delta\theta[\delta]^{12}
-\mbox{$\frac{1}{2}$}\theta[{}^{00}_{00}]^{4}\!\sum_\delta\theta[\delta]^{8})
=0.
$$
There is a unique solution (up to scalar multiple):
$$
(b_1,b_2,b_3,b_4)\,=\,\mu(4,4,-3,-12)\qquad(\mu\in\CC).
$$
With $\mu=1$ and the formula for $\Xi_6[{}^{00}_{00}]$ from section
\ref{g=2} one finds:
{\renewcommand{\arraystretch}{1.5}
$$
\begin{array}{rl}
\Bigl(\theta[{}^{000}_{000}]^4(4F_1+4F_2-3F_3)-12G[{}^{000}_{000}]
\Bigr)(\tau_{1,2})&=\\
\Bigl(\theta[{}^0_0]^4\eta^{12}\Bigr)(\tau_1)
\Bigl(\theta[{}^{00}_{00}]^4
(8\theta[{}^{00}_{00}]^{12}+4\sum_\delta\theta[\delta]^{12}
-6\theta[{}^{00}_{00}]^{4}\sum_\delta\theta[\delta]^{8})\Bigr)(\tau_2)
&=\\
12\Bigl(\theta[{}^0_0]^4\eta^{12}\Bigr)(\tau_1)
\Bigl(\theta[{}^{00}_{00}]^4\Xi_6[{}^{00}_{00}]\Bigr)(\tau_2).&
\end{array}
$$
}
Hence the modular form $\Xi_8[{}^{000}_{000}]$, of weight $8$ on $\Gamma_3(1,2)$ defined by
$$
\Xi_8[{}^{000}_{000}]\,:=\,
\bigl(\theta[{}^{000}_{000}]^4(4F_1+4F_2-3F_3)-12G[{}^{000}_{000}]
\bigr)/12
$$
satisfies all the constraints except maybe (iii$_0$)(2). To check this last constraint, let 
$M\cdot[{}^{000}_{000}]=[{}^{abc}_{def}]$ with $ad=1$, so $a=d=1$.
As $G[{}^{000}_{000}](\tau)=\theta^2[{}^{000}_{000}](\tau)G^\flat[{}^{000}_{000}](\tau)$ for a holomorphic function $G^\flat[{}^{000}_{000}]$,
$G[{}^{abc}_{def}](M\cdot\tau)$ is the product of
$\theta^2[{}^{abc}_{def}](\tau)$ and a holomorphic function, hence
$G[{}^{abc}_{def}](M\cdot\tau_{1,2})=0$ because $\theta^2[{}^{abc}_{def}](\tau_{1,2})=
\theta^2[{}^1_1](\tau_1)\theta^2[{}^{bc}_{ef}](\tau_2)=0$.

We conclude that $\Xi_8[{}^{000}_{000}]$, defined as above, satisfies
all three constraints.

In \cite{DG} we will show
that it is the only modular form of weight $8$ on $\Gamma_3(1,2)$
which satisfies the constraints. This then implies that the desired functions $\Xi_6[\Delta]$ from \cite{DP2} indeed do not exist because $G[{}^{000}_{000}]$
is not the product of $\theta[{}^{000}_{000}]^4$ with a modular form of weight $6$.


\subsection{The cosmological constant}
In supersymmetric string theories one expects for the cosmological constant to vanish because of perfect cancellation between the positive
contribution from bosonic states and the negative one from fermionic states. As a consistency check we will show that our solution for the
chiral measure gives a vanishing contribution to the cosmological constant. Like in \cite{DPg2b}, for type II strings the GSO projections
gives $c_{\Delta, \Delta'}=1$ and we will prove that
$$
\sum_{\Delta} {\rm d}\mu [\Delta]=0,\qquad
\mbox{equivalently}\quad
\Bigl(\sum_\Delta\,{\Xi}_8[\Delta]\Bigr)(\tau)\,=\,0
$$
for all $\tau\in\HH_3$.

The sum of the $36$ functions ${\Xi}_8[\Delta]$
is invariant under $Sp(6,\FF_2)$, hence it is a modular form of weight $8$
on $Sp(6,\ZZ)$.
In \cite{DG} we will show that it must then be a
scalar multiple of $\sum_\Delta\theta[\Delta]^{16}$:
$$
\Bigl(\sum_\Delta\,{\Xi}_8[\Delta]\Bigr)(\tau)\,=\,
\lambda \Bigl(\sum_\Delta \theta[\Delta]^{16}\Bigr)(\tau).
$$
The function $\sum_\Delta\,{\Xi}_8[\Delta]$ is given by:
$$
4\sum_\Delta\theta[\Delta]^{16}
+4\sum_\Delta \theta[\Delta]^4
\Bigl(\sum_{\Delta'}\epsilon_{\Delta,\Delta'}\theta[\Delta']^{12}\Bigr)
-3\Bigl(\sum_\Delta\theta[\Delta]^{8}\Bigr)^2
-12\sum_\Delta G[\Delta],
$$
where the constants $\epsilon_{\Delta,\Delta'}=\pm 1$ are determined by the
transformation theory of the theta constants.

We will show that $\lambda=0$ by taking first $\tau=diag(\tau_1,\tau_2,\tau_3)$
and then let $\tau_1,\tau_2,\tau_3\rightarrow i\infty$.
On the theta constants this gives
$$
\theta[{}^{abc}_{def}]\longmapsto \left\{
\begin{array}{lr}
1&\mbox{if}\;a=b=c=0,\\
0&\mbox{else},
\end{array}\right.
\qquad\mbox{hence}\quad
\left\{
\begin{array}{lr}
\sum_\Delta\theta[\Delta]^{16}&\longmapsto 8,\\
\sum_\Delta\theta[\Delta]^{8}&\longmapsto 8.
\end{array}\right.
$$
In the summand $\sum_\Delta \theta[\Delta]^4
(\sum_{\Delta'}\epsilon_{\Delta,\Delta'}\theta[\Delta']^{12})$
we thus need only consider the terms with $\Delta=[{}^0_b]$,
$\Delta'=[{}^0_{b'}]$.
The terms with $\Delta=[{}^0_b]$
are summands of $\Xi_8[{}^0_b](\tau)$. Let $M$ be the symplectic matrix
$$
M=\begin{pmatrix}I&B\\0&I\end{pmatrix},\quad
B=diag(b_1,b_2,b_3),\qquad\mbox{so}\quad
M\cdot [{}^0_{b'}]=[{}^{\;\;0}_{b+b'}].
$$
In particular, $M\cdot [{}^0_0]=[{}^0_b]$ and thus
$\Xi_8[{}^0_b](\tau)=\Xi_8[{}^0_0](M^{-1}\tau)$ (note that $\gamma(M,M^{-1}\cdot\tau)=1$).
{F}rom the definition of the theta constants as series in \ref{Heisenberg}
it is obvious that
$$
\theta[{}^0_{b'}]^4(M^{-1}\cdot\tau)\,=\,\theta[{}^{\;\;0}_{b+b'}]^4(\tau)
$$
hence $\epsilon_{\Delta,\Delta'}=+1$ if $\Delta=[{}^0_b]$,
$\Delta'=[{}^0_{b'}]$.
Thus we get:
$$
\sum_\Delta \theta[\Delta]^4
(\sum_{\Delta'}\epsilon_{\Delta,\Delta'}\theta[\Delta']^{12})
\,\longmapsto\,\sum_{b}\theta[{}^0_b]^4
(\sum_{b'}\epsilon_{[{}^0_b],[{}^0_{b'}]}\theta[{}^0_{b'}]^{12})
\,\longmapsto\,8\cdot 8=64.
$$
Finally, each $G[\Delta]$ is a sum of $P_L$'s and each $P_L$ is a product
of eight distinct theta constants. Thus all $P_L$'s map to zero except for
$P_{L_0}:=\prod_{d,e,f}\theta[{}^{000}_{def}]$ which maps to $1$.
Note that $L_0=\{\,({}^{abc}_{000})\,\}$ and that $L_0\subset Q_\Delta$
iff $\Delta=[{}^{000}_{def}]$. Thus exactly $8$ of the $G[\Delta]$
map to one, and the others map to zero.
The constant $\lambda$ can now be determined:
$$
4\cdot8+4\cdot 8^2-3\cdot 64-12\cdot 8=\lambda\cdot 8\qquad
\Longrightarrow\quad \lambda=0,
$$
hence the cosmological constant is zero.


\section{A conjecture of D'Hoker and Phong}\label{conjecture}
In \cite{DP1}, section 4.1, D'Hoker and Phong conjecture that:
$$
2^g\Psi_8(\tau)-\Psi_4^2(\tau)\,=\,0\qquad(\forall\tau\in J_g\;(\subset\HH_g))
$$
where $J_g$ is the closure in $\HH_g$ of the set of all period matrices $\tau$ of Riemann surfaces of genus $g$, in all genera $g$, where (\cite{DP1}, (3.10))
$$
\Psi_{4k}(\tau)\,:=\,\sum_\Delta\,\Theta[\Delta]^{8k}(\tau),
$$
the sum is over the $2^{g-1}(2^g+1)$ even characteristics $\Delta$.
They verify this conjecture for $g=1,2$.

The conjecture is actually known to be true for $g\leq 4$, due to
results of Igusa in \cite{IgusaC} which we briefly sketch now.
Schottky discovered  a modular form $J$ of weight $8$ on $\HH_4$,
whose zero locus is known to be $J_4$.
According to \cite{IgusaC}, Theorem 1 and its proof, the
modular form $J$ is, up to a scalar multiple, equal to
$$
(2^{-4}\Psi_4(\tau))^2-2^{-4}\Psi_8(\tau)\,=\,
-2^{-8}(2^4\Psi_8(\tau)-\Psi_4^2(\tau))
$$
(use that Igusa's $s_{0,16}=\Psi_8$, cf.\ his definition on p.\ 356).
This verifies the conjecture for $g=4$.

For the case $g=3$ we already verified the conjecture in section \ref{weight 8}. Alternatively,
one can use the Siegel operator, which is:
$$
\phi(f)(\tau_{g-1})\,:=\,\lim_{\tau_{g}\rightarrow i\infty} f(\tau_{g-1,1})
$$
where $\tau_{g-1,1}$ is the matrix in $\HH_g$ with
diagonal blocks $\tau_{g-1}\in\HH_{g-1}$ and $\tau_{g}\in\HH_1$ (and the other coefficients
are zero).
It is easy to verify that $\phi$ maps $\theta[{}^{abcd}_{efgh}]$ to zero if $d=1$ and else the result
is $\theta[{}^{abc}_{efg}]$. Thus applying the Siegel operator to
$2^4\Psi_8-\Psi_4^2$ we get:
$$
\phi(2^4\Psi_8-\Psi_4^2)(\tau_3)\,=\,(2^5\Psi_8-(2\Psi_4)^2)(\tau_3)
\,=\, 4(2^3\Psi_8-\Psi_4^2)(\tau_3).
$$
As any element in $\HH_3$, viewed as boundary component of $\HH_4$, is in the closure of $J_4$, the modular form $2^3\Psi_8-\Psi_4^2$,
of weight $8$ on $Sp(6,\ZZ)$ is identically zero on $\HH_3$.

\section{Conclusions}
In this paper we considered the problem of finding the chiral measure for supersymmetric strings.
We have taken the stance that the vacuum to vacuum amplitude should split as in (\ref{ampiezgenh}). 
Such a point of view requires a detailed analysis of the geometry of the moduli space of super Riemann surfaces,
however, we did not perform such an analysis here.

\

Instead, we determined the weakest possible constraints that should be obeyed by the measures ${\rm d}\mu[\Delta^{(3)}]$ 
in order to provide a modular invariant expression for the vacuum to vacuum amplitude. 
Our constraints are a slight modification 
of the ones of D'Hoker and Phong in \cite{DP2}. 
Using the transformation properties of certain basic functions under suitable  subgroups of the modular group, we found that it suffices to consider only one fixed spin structure. This led us to consider
modular forms of weight $8$ on $\Gamma_3 (1,2)$ and
we explicitly identified the constraints on these modular forms. 

The strategy for searching solutions has been to use group representation theory on the space of modular forms.
This permitted us to reproduce the genus $g=1,2$ results and to find a solution for the $g=3$ case. Indeed, it happens that
our solution is also unique, but this will be proved in \cite{DG}, 
together with a systematic study of the modular group representation on modular forms on $\Gamma_3(2)$. 
We also checked that our result is consistent with the vanishing of the cosmological constant, and proved a conjecture of D'Hoker and Phong.

\

Although the question of the chiral splitting for superstrings in the genus $g=3$ case is open, we think that our results provide some evidence for a positive answer.

\


\newpage

\section{A: Characteristics and quadrics}
\label{A}\label{quasymp}
We recall the basics about characteristics,
in particular their relation to quadratic forms on $V=\FF_2^{2g}$,
where $\FF_2:=\ZZ/2\ZZ$ is the field with two elements
(cf.\ \cite{Igusa}, $\S$ 5.6, but we use the additive convention).
We introduce a symplectic form on $V$ (so $E$ is non-degenerate, bilinear and $E(v,v)=0$ for all $v\in V$),
$$
E:V\times V\, \longrightarrow\, \FF_2,\qquad
E(v,w):=v_1w_{g+1}+v_2w_{g+2}+\ldots+v_gw_{2g}+
v_{g+1}w_1+\ldots+v_{2g}w_g.
$$
We consider the quadratic forms $q$ on $V$ whose associated bilinear form is $E$, that is the maps
$$
q:\,V\,\longrightarrow\,\FF_2,\qquad q(v+w)=q(v)+q(w)+E(v,w).
$$
Note that $q(ax)=a^2q(x)=aq(x)$ for $a\in \FF_2$ i.e.\ $a=0,1$.
If $q,q'$ are such quadratic forms then, as $2=0$ in $\FF_2$,
$q+q'$ (defined as usual: $(q+q')(v)=q(v)+q'(v)$) is linear in
$v$: $(q+q')(v+w)=(q+q')(v)+(q+q')(w)$ and hence there is a $w\in V$
such that $q'(v)=q(v)+E(v,w)$ for all $v\in V$. Conversely, if $E$ is
associated to $q$ and $w\in V$,
then $q'$, defined by $q'(v)=q(v)+E(v,w)$ also has $E$ as associated bilinear form.
Thus, once we fix $q$, for each of the $2^{2g}$ elements $w\in V$ we have obtained
a quadratic form
whose associated bilinear form is $E$ and all quadratic forms
associated to $E$ are obtained in this way.
One verifies easily that for all $\epsilon_i,\epsilon_i'\in\FF_2$
the function
$$
q(v)=v_1v_{g+1}+v_2v_{g+2}+\ldots+v_gv_{2g}+\epsilon_1v_1+\ldots+
\epsilon_gv_g+\epsilon_1'v_{g+1}+\ldots+\epsilon_g'v_{2g}
$$
satisfies $q(v+w)=q(v)+q(w)+E(v,w)$. In this way we obtain $2^{2g}$
quadratic forms associated to $E$, and thus
each of the $2^{2g}$ quadratic forms associated to $E$
is defined by certain $\epsilon_i,\epsilon_i'$, $i=1,\ldots,g$.
The characteristic associated to $q$ is defined as
$$
\Delta_q:=\left[{}^{\epsilon_1\,\epsilon_2\,\ldots\,\epsilon_g}_
{\epsilon_1'\,\epsilon_2'\,\ldots\,\epsilon_g'}\right],
\qquad \mbox{let}\quad
e(\Delta_q):=(-1)^{\sum_{i=1}^g \epsilon_i\epsilon'_i}\quad(\in\{1,-1\}).
$$
We say that $\Delta_q$ (or $q$) is even if $e(\Delta_q)=+1$ and odd else.
One can verify that
$$
e(\Delta_q)2^g\,=\,\sum_{v\in V} (-1)^{q(v)}.
$$
For example in case $g=1$ and $q(v)=v_1v_2$ then
$q(v)=0$ for $v=(0,0),\,(0,1),\,(1,0)$ and $q(v)=1$ for $v=(1,1)$
so $2e([{}^0_0])=3-1=2$; if $q(v)=v_1v_2+v_1+v_2$ then $q(v)=1$ except if $v=(0,0)$ so $2e([{}^1_1])=1-3=-2$.
It follows that $q(v)$ has $2^{g-1}(2^g+1)$ zeroes in $V$ if
$\Delta_q$ is even  and has $2^{g-1}(2^g-1)$ zeroes if $\Delta_q$ is
odd.
Moreover, there are $2^{g-1}(2^g+1)$ even characteristics and
$2^{g-1}(2^g-1)$ odd characteristics.

The group $Sp(2g,\ZZ)$ acts on $V=\ZZ^{2g}/(2\ZZ)^{2g}$, the subgroup
$\Gamma_g(2)$ (see \ref{remark on ii}) acts trivially, so we get an action of the quotient
$Sp(2g,\ZZ)/\Gamma_g(2)\cong Sp(2g,\FF_2)$
(cf.\ \cite{Igusa}, V.6, Lemma 25) on $V$.
Let $q:V\rightarrow \FF_2$ be a quadratic form associated to $E$,
then we define a function $\sigma\cdot q$ on $V$ by
$$
(\sigma\cdot q)(v)\,:=\,q(\sigma^{-1}v)\qquad
(v\in V,\;\sigma\in Sp(2g,\ZZ)).
$$
As $E(\sigma^{-1} v,\sigma^{-1} w)=E(v,w)$ for $\sigma\in Sp(2g,\ZZ)$
and $v,w\in V$, one verifies that also $\sigma\cdot q$ is a quadratic form associated to $E$. Obviously $\sigma\cdot q$ and $q$ have the same number of zeroes, so the parity of $q$ and $\sigma\cdot q$ are the same: $e(\sigma\cdot q)=e(q)$. The group $Sp(2g,\ZZ)$ acts
transitively on the even and the odd quadrics (cf.\ \cite{Igusa},
V.6, Proposition 3). If $\Delta_q=[{}^\epsilon_{\epsilon'}]$ and
we write $v=(v',v'')$ as a row vector with $v',v''\in\FF_2^g$,
then $q(v)=v'{}^tv''+\epsilon {}^tv'+\epsilon'{}^tv''$. With the formula for $\sigma^{-1}$ from section \ref{trans theta4}
one then easily verifies  that
$$
\Delta_{\sigma\cdot q}\,\equiv\,\sigma\cdot\Delta_q\quad
\mbox{mod}\;2
$$
with $\sigma\cdot\Delta_q$ as in section \ref{constraints}.
Thus the interpretation of characteristics as parameters for quadrics
associated to $E$ leads to a transformation formula which is exactly the one of the characteristics of the theta constants
$\theta[\Delta]$ when we consider the $\Delta$ modulo two.

A subspace $W\subset V$ is isotropic if $E(w,w')=0$ for all
$w,w'\in W$. Given a basis $e_1,\ldots,e_k$ of $W$ it is not hard to see that one can extend it to a symplectic basis $e_1,\ldots e_{2g}$
of $V$ (so $E(e_i,e_j)=0$ unless $|i-j|=g$ and then $E(e_i,e_j)=1$).
In particular, the group $Sp(2g,\ZZ)$ acts transitively on the
isotropic subspaces of $V$ of a given dimension. The number of $k$-dimensional isotropic subspaces  of $V\cong \FF_2^{2g}$
is given by
{\renewcommand{\arraystretch}{2}
$$
\begin{array}{rcl}
\frac{(2^{2g}-1)(2^{2g-1}-2)(2^{2g-4}-4)\ldots(2^{2g-(k-1)}-2^{k-1})}
{\sharp GL(k,\FF_2)}&=&
\frac{(2^{2g}-1)(2^{2g-1}-2)(2^{2g-2}-4)\ldots(2^{2g-(k-1)}-2^{k-1})}
{(2^k-1)(2^k-2)\ldots (2^k-2^{k-1})}\\
&=&
\frac{(2^{2g}-1)(2^{2g-2}-1)(2^{2g-4}-1)(2^{2g-6}-1)\ldots(2^{2(g-k)+2}-1)}{
(2^k-1)(2^{k-1}-1)\ldots (2-1)},
\end{array}
$$
} 
in the numerator we count the ordered $k$-tuples of independent
elements $v_1,\ldots,v_k\in V$ with $E(v_i,v_j)=0$ for all $i,j$:
for $v_1$ we can take any element in $V-\{0\}$, for $v_2$ we can take any element in $\langle v_1\rangle^\perp\cong \FF_2^{2g-1}$ except $0,v_1$, so $v_2\in \langle v_1\rangle^\perp-\langle v_1\rangle$, next
$v_3\in\langle v_1,v_2\rangle^\perp-\langle v_1,v_2\rangle$.

If $W_1,\ldots,W_N$ are the $k$-dimensional isotropic subspaces
in an even quadric $Q\subset V$ defined by $q=0$, then
$\sigma(W_1),\ldots,\sigma(W_N)$ are the $k$-dimensional isotropic subspaces in the even quadric $\sigma(Q)\subset V$ defined by
$\sigma\cdot q=0$, indeed $(\sigma\cdot q)(\sigma v)=q(v)$.
In particular, any even quadric in $V$ contains the same number
of isotropic subspaces of a given dimension.
An even quadric contains a maximal isotropic subspace $L$, for example
$L_0=\{({}^{v_1\ldots v_g}_{0\,\ldots\,0}):v_i\in\FF_2\}$ is contained
in the even quadric $Q$ corresponding to the characteristic
$[{}^{0\ldots0}_{0\ldots0}]$. An odd quadric does not contain a
maximal isotropic subspace however: if $L\subset Q$ were such a subspace, then $\sigma(L)=L_0$ for a suitable $\sigma\in Sp(2g,\ZZ)$, if $\sigma( Q)$  corresponds to the characteristic
$[{}^\epsilon_{\epsilon'}]$ then $L_0\subset \sigma(Q)$ implies
that $\epsilon_1=\ldots=\epsilon_g=0$, hence the characteristic must be even. On the other hand, an odd quadric does contain an isotropic subspace of dimension $g-1$, for example
$W_0=\{({}^{v_1\ldots v_{g-1}0}_{\;0\;\ldots\;0\quad0}):v_i\in\FF_2\}$
is contained in the odd quadric with characteristic $[{}^{0\ldots01}_{0\ldots01}]$.

The number of even quadrics which contain a fixed $k$-dimensional isotropic subspace
is easy to count: we may assume that the subspace has basis $e_1,\ldots, e_k$ and then the characteristic of an even quadric
containing it is
$$
\left[\begin{array}{lcr}
0\quad\ \ldots&\!0\quad\epsilon_{k+1}\!&\ldots\quad\epsilon_g\\
\epsilon'_1\quad\ldots&\epsilon'_k\quad\epsilon'_{k+1}&\ldots\quad\epsilon'_g\\
\end{array}\right]
\qquad\mbox{with}\quad
\sum_{i=k+1}^g\epsilon_{k+i}\epsilon'_{k+i}\,=0,
$$
so one has $2^k\cdot2^{g-k-1}(2^{g-k}+1)$ such even quadrics.
To find the number of $k$-dimensional isotropic subspaces in an
even quadric one can now count the pairs $(W,Q)$ of such a subspace $W$
contained in even quadric $Q$ in two ways: first as the product of
the number of $W$ with the number of even $Q$ containing a fixed $W$
and second as the product of the number of even quadrics with
the number of $k$-dimensional isotropic subspaces in an
even quadric. For example the number of pairs $(W,Q)$ of a maximally
isotropic subspace in an even quadric in $\FF_2^6$ is $135\cdot 2^3$,
and thus the number of such subspaces in a fixed $Q$ is $135\cdot 2^3/36=30$.

For small $g$ we list some of these dimensions in the table on the left below,
in the table on the right we list the number of $k$-dimensional isotropic
subspaces contained in an even quadric.
$$
\begin{array}{|@{\hspace{6pt}}l||@{\hspace{6pt}}c|@{\hspace{6pt}}c|
@{\hspace{6pt}}c|@{\hspace{6pt}}c|}
\hline g\,\backslash\,\mbox{dimension}&
1&
2&
3&
4
\\\hline \hline
1&3&& &\\ \hline
2&15&15& &\\ \hline
3&63&315&135&\\ \hline
4&255&5355&11475&2295\\ \hline
\end{array}
\qquad
\begin{array}{|@{\hspace{6pt}}l||@{\hspace{6pt}}c|@{\hspace{6pt}}c|
@{\hspace{6pt}}c|@{\hspace{6pt}}c|}
\hline g\,\backslash\,\mbox{dimension}&
1&
2&
3&
4
\\\hline \hline
1&2&& &\\ \hline
2&9&6& &\\ \hline
3&35&105&30&\\ \hline
4&135&1575&2025&270\\ \hline
\end{array}
$$

\section{B: Transformation theory of theta constants}
\label{appendix trans}
\subsection{Transformation formula for the $\theta[\Delta]$}
\label{trans theta}
We recall the transformation formula for the functions $\theta[\Delta]$, for an even characteristic $\Delta$
as given in \cite{Igusa}, V.1, Corollary.

Let $\Delta=[{}^a_b]$ with row vectors $a,b\in\ZZ^g$ and $a_i,b_i\in\{0,1\}$.
We consider the characteristic $m=(m',m'')\in(\RR^g)^2$
given by $m'=a/2,m''=b/2$. Then $\Theta[\Delta](\tau)=\theta_m(\tau)$
and
for $\sigma\in Sp(2g,\ZZ)$ the transformation formula is:
$$
\theta_{\sigma\cdot m}(\sigma\cdot \tau)\,=\,\kappa(\sigma)
e^{2\pi i\phi_m(\sigma)}\gamma(\sigma,\tau)^{1/2}\theta_m(\tau),
\qquad \sigma\,=\,\begin{pmatrix}A&B\\C&D\end{pmatrix}\in Sp(2g,\ZZ),
$$
where $\kappa(\sigma)$ is an eight root of unity (\cite{Igusa}, V.3, Theorem 3), $\gamma$ is as in section \ref{basic defs} and
$$
\phi_{m}(\sigma)\,=\,
\sum_{k,l=1}^g \mbox{$\frac{-1}{8}$}
\Bigl(({}^tDB)_{kl}a^{}_ka^{}_l
-2({}^tBC)_{kl}a^{}_kb^{}_l
+({}^tCA)_{kl}b^{}_kb^{}_l\Bigr)
+
\mbox{$\frac{1}{4}$}(({}^tD)_{kl}a_k-({}^tC)_{kl}b_k)
(A{}^tB)_{ll}.
$$

\subsection{Transformation formula for the $\theta[\Delta]^4$}
\label{trans theta4}
We apply this formula above to $\theta_m^4$, so
we get the factor $\kappa(\sigma)^4exp(8\pi i\phi_m(\sigma))\gamma(\sigma,\tau)$.  As $exp(2\pi i n)=1$ for integers $n$, we obtain:
$$
e^{8\pi i\phi_m(\sigma)}\,=\,(-1)^{a{}^tDB{}^ta+b{}^tCA{}^tb}.
$$

The condition that $\sigma\in Sp(2g,\ZZ)$ is that
$\sigma E{}^t\sigma =E$ where $E$ has blocks $A=D=0,B=-C=I$:
$$
\sigma E{}^t\sigma =E\qquad\mbox{iff}\quad
\begin{pmatrix}-B{}^tA+A{}^tB&-B{}^tC+A{}^tD\\
-D{}^tA+C{}^tB&-D{}^tC+C{}^tD\end{pmatrix}=
\begin{pmatrix}0&I\\-I&0\end{pmatrix}.
$$
As $E^{-1}=-E$ we find
that
$$
\sigma^{-1}=-E{}^t\sigma E,
\qquad \sigma^{-1}=\begin{pmatrix}{}^tD&-{}^tB\\-{}^tC&{}^tA\end{pmatrix}.
$$
As $\sigma\in Sp(2g,\ZZ)$, also $\sigma^{-1}\in Sp(2g,\ZZ)$,
thus $\sigma$ satisfies also
${}^tBD-{}^tDB=0$ and ${}^tAC-{}^tCA=0$,
that is, ${}^tDB$ and ${}^tCA$ are symmetric matrices.
Hence the integers $a_ka_l,b_kb_l$ in
$a{}^tDB{}^ta+bCA{}^tb$ are multiplied by an even integer if $k\neq l$,
and thus they do not contribute to $e^{8\pi i\phi_m(\sigma)}$.
In the exponent there remains
$\sum_k(a_k^2({}^tDB)_{kk}+b_k^2({}^tCA)_{kk})$, but note that $a_k^2\equiv a_k$ mod $2$.
For a $g\times g$ matrix $M$, let $\diag(M)$ be the column vector $(M_{11},M_{22},\ldots,M_{gg})$
of diagonal entries. Then we get the formula:
$$
e^{8\pi i\phi_m(\sigma)}\,=\,
(-1)^{a\diag({}^tDB)+b\diag({}^tCA)}.
$$

Next we consider the case that $\sigma\in\Gamma_g(1,2)$. Then also
$\sigma^{-1}\in \Gamma_g(1,2)$ which implies that the diagonals of
${}^tDB$ and ${}^tCA$ are zero mod $2$, hence
we conclude that $e^{8\pi i\phi_m(\sigma)}=1$
for all $\sigma\in\Gamma_g(1,2)$.

A final remark is that $\sigma\cdot m$ in \cite{Igusa} is computed in $\RR^{2g}$ whereas we normalize the characteristics modulo vectors in $\ZZ^{2g}$ to have coefficients
$m'_i,m''_i\in\{0,1/2\}$. This is justified for the $\theta_m^4$ by formula
($\theta$.2) in \cite{Igusa}, I.10.

\subsection{Applications}\label{applications}
Let $[{}^0_0]=[{}^{0\ldots0}_{0\ldots0}]$ be a genus $g$ characteristic and let
$$
f_1:=\theta[{}^{0}_{0}]^{12},\qquad
f_2:=\sum_{\Delta^{(g)}}\theta[\Delta^{(g)}]^{12},\qquad
f_3:=\theta[{}^{0}_{0}]^4\sum_{\Delta^{(g)}}\theta[\Delta^{(g)}]^{8},
$$
where we sum over the even characteristics $\Delta^{(g)}$ in genus $g$.
We show that the functions $\theta[{}^{0}_{0}]^4f_i$, $i=1,2,3$, are modular forms of weight $8$ for $\Gamma_g(1,2)$.

In the cases $i=1,3$ these functions are polynomials of degree two in the $\theta[\Delta^{(g)}]^8$.
The transformation formula for $\theta_m^8$ (note $\kappa(\sigma)^8=1$, $e^{16\pi i\phi_m(\sigma)}=1$), is in our notation:
$$
\theta[\Delta^{(g)}]^8(\sigma\cdot \tau)\,=\,\gamma(\sigma,\tau)^4
\theta[\sigma^{-1}\cdot\Delta^{(g)}]^8(\tau)\qquad(\sigma\in Sp(2g,\ZZ)).
$$
Hence the $\theta[\Delta]^8$ are permuted by the action of $\sigma\in Sp(2g,\ZZ)$,
up to the common cocycle $\gamma(\sigma,\tau)^4$. In particular, if $\sigma\in \Gamma_g(1,2)$ then $\sigma^{-1}\cdot [{}^0_0]=[{}^0_0]$ and it follows that the
$\theta[{}^{0}_{0}]^4f_i$, $i=1,3$, are modular forms of weight $8$ for $\Gamma_g(1,2)$.

In case $i=2$ we use the formula for the $\theta_m^4$. For
$\sigma\in \Gamma_g(1,2)$ we have $ e^{8\pi i\phi_m(\sigma)}=1$
so in our notation we get:
$$
\theta[\Delta^{(g)}]^4(\sigma\cdot \tau)\,=\,\kappa(\sigma)^4\gamma(\sigma,\tau)^2
\theta[\sigma^{-1}\cdot\Delta^{(g)}]^4(\tau)\qquad(\sigma\in\Gamma_g(1,2)).
$$
Hence the $\theta[\Delta^{(g)}]^4$ are permuted by the action of
$\sigma\in \Gamma_g(1,2)$ up to a common (i.e.\ independent of
$\Delta^{(g)}$) factor $\kappa(\sigma)^4\gamma(\sigma,\tau)^2$ and these $\sigma$ fix $[{}^0_0]$. Thus $\theta[{}^{0}_{0}]^4f_2$ transforms with the factor
$\kappa(\sigma)^{16}\gamma(\sigma,\tau)^8$, but as $\kappa(\sigma)^8=1$ for any $\sigma\in Sp(2g,\ZZ)$, this implies that it is a modular form of weight $8$ on
$\Gamma_g(1,2)$.

\subsection{Transformation formula for the $P_L$}\label{trans PL}
In section \ref{Gdelta} we defined, for a Lagrangian subspace
$L$ of $V=\FF_2^6$ the function
$
P_L=\prod_{Q\supset L}\theta[\Delta_Q]^2
$
where the product is over the eight even quadrics which contain $L$.
We will show that these functions are permuted, up to a factor $\gamma(\sigma,\tau)^8$, by the action of $\sigma\in Sp(6,\ZZ)$.

We write representatives in $\ZZ^6$ for the eight elements of $L$ as
$$
L\,=\,\left\{
\,p^{(0)}=\left(
{}^{x^{(0)}_1x^{(0)}_2x^{(0)}_3}_{y^{(0)}_1y^{(0)}_2y^{(0)}_3}
\right),
\ldots,
p^{(7)}=\left(
{}^{x^{(7)}_1x^{(7)}_2x^{(7)}_3}_{y^{(7)}_1y^{(7)}_2y^{(7)}_3}
\right)\,
\right\},\qquad
x^{(j)}_k,y^{(j)}_l\in\{0,1\},
$$
and we will assume that $p^{(0)}=({}^0_0)$.
As $L$ is a subgroup of $\FF_2^6$, it is not hard to see that
$$
\sum_{j=0}^7 x^{(j)}_k\,\equiv\,\sum_{j=0}^7 y^{(j)}_k\,\equiv\,0
\;\mbox{mod}\;4,\qquad
\sum_{j=0}^7x_k^{(j)}x_l^{(j)}\,\equiv\,\sum_{j=0}^7y_k^{(j)}y_l^{(j)}
\,\equiv\,\sum_{j=0}^7x_k^{(j)}y_l^{(j)}
\equiv\,0\;\mbox{mod}\;2
$$
for $k,l=1,2,3$ (use for example $({}^x_y)\mapsto x_k$ (or $y_k)$
is a homomorphism of $L\cong\FF_2^3$ to $\FF_2$ and thus
each fiber has either $4$ or $8$ elements; similarly,
the fibers of the homomorphism
$L\rightarrow \FF_2^2$, $({}^x_y)\mapsto (x_k,x_l)$
contain an even number of elements etc.).

Let $Q_{\Delta}$  be one of the quadrics
containing $L$ with $\Delta=[{}^a_b]$.
Then $\Delta^{(j)}:=[{}^{a+y^{(j)}}_{b+x^{(j)}}]$
is an even characteristic:
$$
(a+y^{(j)}){}^t(b+x^{(j)})=
a{}^tb+x^{(j)}{}^ty^{(j)}+a{}^tx^{(j)}+b{}^ty^{(j)}\equiv\,
0+q_\Delta(p^{(j)})\quad\mbox{mod}\;2,
$$
with $q_\Delta$ the quadratic form defining $Q_\Delta$; 
as $L\subset Q_\Delta$ we have $q_\Delta(p^{(j)})=0$
for all $j$ and thus $[{}^{a+y^{(j)}}_{b+x^{(j)}}]$ is indeed even.
Moreover, $L\subset Q_{\Delta^{(j)}}$ because
$$
q_{\Delta^{(j)}}(p^{(k)})=
x^{(k)}{}^ty^{(k)}+(a+y^{(j)}){}^tx^{(k)}+(b+x^{(j)}){}^ty^{(k)}
\equiv\, q_\Delta(p^{(k)})+E(p^{(j)},p^{(k)})
\equiv0 \quad\mbox{mod}\;2.
$$

In this way, given $\Delta$, we get $7$ other even characteristics
$\Delta^{(j)}$ of quadrics which contain $L$.
Therefore the characteristics of the eight even $Q$ with $Q\supset L$ are:
$$
[{}^{a^{(j)}}_{b^{(j)}}],\qquad
a^{(j)}\equiv a+y^{(j)}\;\mbox{mod}\;2,
\quad
b^{(j)}\equiv b+x^{(j)}\;\mbox{mod}\;2,\qquad
a^{(j)},b^{(j)}\in \{0,1\}.
$$
It follows that we have the following
congruences for the coefficients of the characteristics:
$$
\sum_{j=0}^7 a^{(j)}_k\,\equiv\,\sum_{j=0}^7 b^{(j)}_k\,\equiv\,0
\;\mbox{mod}\;4,\qquad
\sum_{j=0}^7a_k^{(j)}a_l^{(j)}\,\equiv\,\sum_{j=0}^7b_k^{(j)}b_l^{(j)}
\,\equiv\,\sum_{j=0}^7a_k^{(j)}b_l^{(j)}
\equiv\,0\;\mbox{mod}\;2.
$$

The transformation formula for the theta constants, given
in section \ref{trans theta}, shows that
$$
P_{\sigma \cdot L}(\sigma\cdot \tau)\,=\,
\kappa(\sigma)^{16}
e^{4\pi i\sum_{j=0}^7\phi_{m_j}(\sigma)}
\gamma(\sigma,\tau)^8P_L(\tau)
$$
where we wrote $m_j:=(a^{(j)},b^{(j)})/2$.
As $\kappa(\sigma)$ is an eight root of unity we get $\kappa(\sigma)^{16}=1$. As we observed in section \ref{trans theta4}, the matrices
${}^tDB$ and ${}^tCA$ are symmetric.
The first term in $2\sum_j\phi_{m_j}(\sigma)$ is then
$$
\mbox{$\frac{-1}{4}$}\sum_{j=0}^7\sum_{k,l=1}^3 ({}^tDB)_{kl}a^{(j)}_ka^{(j)}_l\,=\,
\mbox{$\frac{-1}{4}$}\sum_{k}^3 ({}^tDB)_{kk}
(\sum_{j=0}^7a^{(j)}_k)-
\mbox{$\frac{1}{2}$}\sum_{k<l}^3({}^tDB)_{kl}
(\sum_{j=0}^7a^{(j)}_ka^{(j)}_l),
$$
hence this is an integer. Similarly the third term (with $({}^tCA)$)
is an integer. The second term
$$
\mbox{$\frac{-1}{2}$}\sum_{j=0}^7\sum_{k,l=1}^3 ({}^tBC)_{kl}a^{(j)}_kb^{(j)}_l\,=\,
\mbox{$\frac{-1}{2}$}\sum_{k,l=1}^3({}^tBC)_{kl}
(\sum_{j=0}^7a^{(j)}_kb^{(j)}_l)
$$
is an integer because $\sum_ja^{(j)}_kb^{(j)}_l$ is even for all $k,l$.
The last term is also an integer because it is linear in each $a_k,b_l$
and $\sum_ja_k^{(j)}\equiv \sum_jb_k^{(j)}\equiv 0$ mod $4$.

Again we observe that $\sigma\cdot m$ in \cite{Igusa} is computed in $\RR^{2g}$ whereas we normalize the characteristics modulo vectors in $\ZZ^{2g}$ to have coefficients
$m'_i,m''_i\in\{0,1/2\}$. This is justified for the $\theta_m^2$ by formula
($\theta$.2) in \cite{Igusa}, I.10.

Thus we showed that $e^{4\pi i\sum_{j=0}^7\phi_{m_j}(\sigma)}=1$ for all $\sigma$ and it follows that
$P_{\sigma (L)}(\sigma\cdot \tau)=\gamma(\sigma,\tau)^8P_L(\tau)$,
as desired.

\


\section{C: The restriction of the $G[\Delta]$ to $\Delta_{1,2}$}
\label{res12}
\subsection{The restriction of the $P_L$'s to the diagonal.}
\label{res PL}
Now we determine, for each Lagrangian subspace $L\subset V$,
the functions
$P_L(\tau_{1,2})$
which are modular forms in both $\tau_1$ and $\tau_2$.

We already recalled that $\theta[{}^{abc}_{def}](\tau_{1,2})=
\theta[{}^a_b](\tau_1)\theta[{}^{bc}_{ef}](\tau_2)$. This `decomposition' of the characteristic $\Delta=[{}^{abc}_{def}]$
corresponds to the restriction of the quadratic form $q_\Delta$
to the two summands in:
$$
V\,=\,l\,\oplus \,l^\perp,\qquad
l\,:=\,\{\,\left({}^{000}_{000}\right),\;
\left({}^{100}_{000}\right),\;\left({}^{000}_{100}\right),\;
\left({}^{100}_{100}\right)\,\}
$$
so $l$ is a two dimensional symplectic subspace of $V$
(that is, the restriction of the symplectic form $E$ to $l\times l$ is non-degenerate) and its
perpendicular is the four dimensional symplectic subspace
$$
l^\perp\,:=\,\{v\in V:\,E(v,w)=0\;\forall w\in l\,\}\,=\,
\{\left({}^{0ab}_{0cd}\right)\in V\,:\,a,\ldots,d\in\FF_2\,\}.
$$
The restriction of $q_\Delta$ to $l$ (resp.\ $l^\perp$)
is the quadratic form
on $\FF_2^2$ (resp.\ on $\FF_2^4$), associated to the characteristic $[{}^a_d]$ (resp.\ $[{}^{bc}_{ef}]$):
$$
q_{[{}^a_d]}({v_1},{v_4})=v_1v_4+av_1+dv_4,\quad
q_{[{}^{bc}_{ef}]}({v_2,v_3},{v_5,v_6})=v_2v_5+v_3v_6+
bv_2+cv_3+ev_5+fv_6,
$$
and $q_\Delta=q_{[{}^a_d]}+q_{[{}^{bc}_{ef}]}$.

Let $L$ be a Langrangian subspace in $V$ and consider
the intersection $L\cap l$. As $E$ is non-degenerate on $l$, but is
identically zero on $L$, we cannot have $l\subset L$. Thus
$\dim L\cap l\leq 1$.
For dimension reasons, $\dim L\cap l^\perp\geq 1$ and it is at most two since $L\cap l^\perp$ is an isotropic subspace of $l^\perp$. We will
show that
$$
\dim L\cap l\,=0\;\Longleftrightarrow \dim L\cap l^\perp\,=1,\qquad
\dim L\cap l\,=1\;\Longleftrightarrow \dim L\cap l^\perp\,=2.
$$
An example of the first case is
$$
L=\langle\,({}^{000}_{001}),\;({}^{000}_{110}),\;({}^{110}_{000})\,
\rangle,\qquad
L\cap l=\{0\},\quad
L\cap l^\perp=\langle ({}^{000}_{001})\rangle,
$$
whereas $L_0=\{({}^{abc}_{000})\}$ is an example of the second case.

To prove the assertions, consider the exact sequence
$$
0\longrightarrow L\longrightarrow V\stackrel{\phi}{\longrightarrow}\, L^*=\mbox{Hom}(L,\FF_2)\longrightarrow 0,\qquad
\phi(v)(v'):=E(v,v'),
$$
($v\in V,v'\in L$),
note that
$\ker(\phi)=L$ because so $E(v,v')=0$ for all $v'\in L$ implies
$v\in L$ by maximality of $L$, hence $\dim\im(\phi)=6-3=3=\dim L^*$.

In case $L\cap l^\perp$ is one dimensional, that is, $\dim(\ker(\phi)\cap l^\perp)=1$, the subspace $\phi(l^\perp)$
is three dimensional
so $\phi(l^\perp)=L^*$.
Hence for any non-zero $v'\in L$ there is a $v\in l^\perp$ with
$E(v,v')\neq 0$, and thus $L\cap l=\{0\}$.
In case $L\cap l^\perp$ is two dimensional,
$\phi(l^\perp)$ is also two dimensional and hence there is a
$w\in L-\{0\}$ such that $E(v,w)=0$ for all $v\in l^\perp$.
Hence $w\in l$, so in this case
$L\cap l=\{0,w\}$. This concludes the proofs of the assertions.

In the first case we claim that there exists an even $q$
such that $L\subset (q=0)$ and such that the restriction of $q$
to $l$ is the odd quadratic form (with characteristic $[{}^1_1]$),
hence $P_L$ is identically zero on the diagonal.
To see this,
let $w_0$ be the unique non-zero element in $l^\perp\cap L$.
Any $w\in L$ can be written uniquely as $w=w_l+w_p$ with
$w_l\in l$ and $w_p\in l^\perp$. As $L$ is isotropic, we have
$0=E(w_0,w)=E(w_0,w_p)+E(w_0,w_l)$, but $E(w_0,w_l)=0$ as
$w_0\in l^\perp$ and $w_l\in l$, hence $E(w_0,w_p)=0$.
Let $q'$ be an odd quadratic form on $l^\perp$ which is zero in $w_0$: $q'(w_0)=0$. Then
$$
q'(w_0+w_p)\,=\,q'(w_0)+q'(w_p)+E(w_0,w_p)\,=\,q'(w_p).
$$
In case $w_p\neq 0,w_0$, we cannot have $ q'(w_p)=0$, since
then $q'=0$ would contain the maximal isotropic subspace
$\langle w_0,w_p\rangle$ of $l^\perp$, but odd quadrics do not contain maximal isotropic subspaces. Hence $q'(w_p)=q'(w_0+w_p)=1$
for any $w\in L$.
Now let $q''$ be the unique odd quadratic form on $l$ and define
$$
q:\,l\oplus l^\perp\,=V\,\longrightarrow \,\FF_2,\qquad
q(l_0,l_p):=q''(l_0)+q'(l_p).
$$
Then $q$ is an even quadratic form on $V$ which restricts
to the odd quadratic form $q''$ on $l$.
Moreover, $L\subset(q=0)$ because if $w\in L$
then either $w=0$ and so obviously $q(0)=0$, or $w=w_0\in l^\perp$ and $q(w_0)=q'(w_0)=0$ or $w=w_l+w_p$ with $w_l\neq 0$
and $w_p\neq 0$, so $q''(w_l)=1$ and $q'(w_p)=1$ hence  $q(w)=0$.
(In the example above one has
$L\cap l^\perp=\langle ({}^{000}_{001})\rangle$ and then
$q'=q_{[{}^{1f}_{10}]}$, with $f=0,1$ and thus $q=q_\Delta$ with $\Delta=[{}^{11f}_{110}]$.)
Thus the summands $P_L$ of $G[\Delta]$ such that $L\cap l=\{0\}$
are identically zero on the diagonal.

It remains to consider those $L$ such that $ L\cap l=\{0,w\}$ for a unique non-zero $w=w_L\in l$.
In that case $L\cap l^\perp$ is a maximal ($2$-dimensional)
isotropic subspace $L_0$ of $l^\perp$. If $L\subset Q$,
the restriction of $Q$ to $l^\perp$ contains $L_0$. There
are four even quadratic forms $q'$ on $l^\perp$ containing a maximal isotropic subspace, let ${\delta}_1,\ldots,{\delta}_4$ be their characteristics.
There are two even quadratic forms $q''$ on $l$ with $q''(w)=0$, let $\bar{\delta}_1,\bar{\delta}_2$ be their characteristics.
Defining, as before, $q=q''+q'$ we get $2\cdot 4=8$ even quadrics which contain $L$. As $L$ is contained in exactly eight quadrics,
this implies that these are exactly the quadrics containing $L$
and the product of the squares of the corresponding theta nulls
is $P_L$.
This implies that $P_L$ restricts to
$$
P_L(\tau_{1,2})\,=\,
(\theta[{\bar{\delta}_1}]^8\theta[{\bar{\delta}_2}]^8)(\tau_1)
(\theta[{\delta_1}]^4\theta[{\delta_2}]^4
\theta[{\delta_3}]^4\theta[{\delta_4}]^4)(\tau_2).
$$

\subsection{The restriction of the $G[\Delta]$'s to the diagonal.}
\label{res GDelta}
Now we determine, for each even $\Delta$, the functions
$G[\Delta](\tau_{1,2})$
which are modular forms, of weight $8$, in both $\tau_1$ and $\tau_2$.

Recall that $G[\Delta]$ is a multiple of $\theta[\Delta]^2$.
In particular, if
$\Delta\,=\,[{}^{abc}_{def}]$
and $ad=1$, then $\theta[\Delta]$ restricts to zero on
the diagonal $\HH_1\times \HH_2$ and thus
also $G[\Delta]$ restricts to zero.

As we saw in \ref{res PL}, the restriction of $P_L$ to $\HH_1\times\HH_2$
is non-zero iff $L\cap l=\{0,w\}$ for a non-zero $w\in l$.
In that case $L\cap l^\perp=\{0,w_p,w'_p,w_p+w_p'\}$ for some $w_p,w_p'\in l^\perp$ is a maximal isotropic
subspace of $l^\perp$. In particular, $L=(L\cap l)\oplus (L\cap l^\perp)$.
Thus these Lagrangian subspaces correspond to pairs of a non-zero
point in $Q\cap l$ and a Lagrangian plane in $Q\cap l^\perp$.

Let $Q_\Delta\subset V$ be the (even) quadric corresponding to $\Delta$
and assume that $ad=0$. Then $Q\cap l$ is an even quadric in $l$ and consists of three points, $0$ and $w_1,w_2\in l$ (for example if $[{}^a_d]=[{}^0_1]$ then $Q\cap l$ is defined by $v_1v_4+v_4=0$
so consists of the points $({}^{v_1}_{v_4})=({}^0_0),\,({}^1_0),\,({}^1_1)$).
The intersection $Q\cap l^\perp$ is an even quadric in $l^\perp$ and has
$10$ zeroes, such a quadric contains $2\cdot 3=6$ Lagrangian subspaces.

As the Lagrangian subspaces $L\subset V$ such that $P_L$ has non-zero restriction to the diagonal correspond to pairs of a non-zero
point in $Q\cap l$ and a Lagrangian plane in $Q\cap l^\perp$, there
are $2\cdot 6=12$ such subspaces in $Q$.

Using the formula for $P_L(\tau_{1,2})$ from section \ref{res PL},
it follows that the restriction of $G[\Delta]$ is given by:
$$
G[\Delta](\tau_{1,2})\,=\,
(\theta[{}^a_d]^8\theta[\bar{\delta_1}]^8+
 \theta[{}^a_d]^8\theta[\bar{\delta_2}]^8)(\tau_1)
\Bigl(\sum_{L'\subset Q_\Delta\cap l^\perp} \,\prod_{Q'\supset L'} \theta[\delta_{Q'}]^4\Bigr)(\tau_2),
$$
where $[{}^a_d],\,[\bar{\delta}_1],\,[\bar{\delta}_2]$ are the three even
characteristics in $g=1$, $L'$ runs over the six Lagrangian subspaces
in $l^\perp$ which are contained in $Q_\Delta$ and $Q'$ runs over the
even quadrics in $l^\perp$ which contain $L'$.
In particular, the restriction of a $G[\Delta]$ is a product of
a modular form of genus one and one of genus two:
$$
G[{}^{abc}_{def}](\tau_{1,2})\,=\,g_1[{}^a_d](\tau_1)g_2[{}^{bc}_{ef}](\tau_2).
$$

\subsection{The restriction of $G[{}^{000}_{000}]$.}\label{resg00}
The modular forms of genus one $g_1[{}^0_0]$ and genus two $g_2[{}^{00}_{00}]$ are as follows.
{\renewcommand{\arraystretch}{1.5}
$$
\begin{array}{rcl}
g_1[{}^0_0]&=&\theta[{}^0_0]^8\theta[{}^0_1]^8+
\theta[{}^0_0]^8\theta[{}^1_0]^8\\
&=&\theta[{}^0_0]^4(\mbox{$\frac{1}{3}$}f_{21}-\eta^{12}),
\end{array}
$$
}
where we used the formulas from section \ref{dec g=1}.
The genus two modular form is:
{\renewcommand{\arraystretch}{1.5}
$$
\begin{array}{rcl}
g[{}^{00}_{00}]&=&\theta[{}^{00}_{00}]^4\Bigl(
(\theta[{}^{00}_{01}]\theta[{}^{00}_{10}]\theta[{}^{00}_{11}])^4+
(\theta[{}^{00}_{01}]\theta[{}^{10}_{00}]\theta[{}^{10}_{01}])^4+
(\theta[{}^{01}_{00}]\theta[{}^{00}_{10}]\theta[{}^{01}_{10}])^4+
(\theta[{}^{00}_{11}]\theta[{}^{11}_{00}]\theta[{}^{11}_{11}])^4+
 \\ &&\qquad\qquad\qquad +
(\theta[{}^{01}_{00}]\theta[{}^{10}_{00}]\theta[{}^{11}_{00}])^4+
(\theta[{}^{01}_{10}]\theta[{}^{10}_{01}]\theta[{}^{11}_{11}])^4
\Bigr)\\
&=:&\theta[{}^{00}_{00}]^4g_2^\flat[{}^{00}_{00}],
\end{array}
$$
}
where $g_2^\flat[{}^{00}_{00}]$ is a genus two modular form
of weight $6$. A computation, using the methods from \cite{CP}, that
is, using the classical theta formula \ref{classical} to write the
$\theta[\delta]^2$'s in terms of the $\Theta[\sigma]$'s, shows that
$$
g_2^\flat[{}^{00}_{00}]\,=\,
 \mbox{$\frac{1}{3}$}\theta[{}^{00}_{00}]^{12}
+\mbox{$\frac{2}{3}$}\sum_\delta\theta[\delta]^{12}
-\mbox{$\frac{1}{2}$}\theta[{}^{00}_{00}]^{4}\sum_\delta\theta[\delta]^{8},
$$
that is, $
g_2^\flat[{}^{00}_{00}]=\frac{1}{3}f_1+\frac{2}{3}f_2-\frac{1}{2}f_3$
with $f_i$ as in section \ref{g=2}.

\

\end{document}